\documentclass[pdflatex,sn-mathphys-num]{sn-jnl}

\usepackage[
    backend=biber,
    citestyle=nature,
  ]{biblatex}
\usepackage{times}

\usepackage{graphicx}
\usepackage{dcolumn}
\usepackage{bm}
\usepackage{bbm}

\usepackage[utf8]{inputenc}
\usepackage[T1]{fontenc}
\usepackage{amsfonts}
\usepackage{amsmath}
\usepackage{mathptmx}
\usepackage{etoolbox}
\usepackage{booktabs}
\usepackage{tabularx}
\usepackage{multirow}
\usepackage{longtable}
\usepackage{array}
\usepackage{multirow}
\usepackage{makecell}
\usepackage{xcolor}
\usepackage{colortbl}

\usepackage{subfig}
\usepackage{tikz}
\usepackage{hyperref}
\usepackage{cleveref}
\usepackage{placeins}
\usepackage{xurl}

\DeclareMathOperator{\median}{median}
\definecolor{transparent}{gray}{1.}
\definecolor{purple_GNN}{HTML}{a4239c}
\definecolor{green_NetSciML}{HTML}{008381}
\definecolor{blue_line}{HTML}{658cb2}

\usepackage[title]{appendix}%

\newcommand{\mse}{\mathrm{mse}}

\newcommand{\mrm}[1]{\mathrm{#1}}
\newcommand{\pfrt}{p_\mrm{frt}}

\crefname{appendix}{Supplemental Materials}{Supplemental Materials}
\Crefname{appendix}{Supplemental Materials}{Supplemental Materials}
\crefname{apptable}{supplemental table}{supplemental tables}
\Crefname{apptable}{Supplemental Table}{Supplemental Tables}

\makeatletter
\def\cline#1{\@cline#1\@nil}
\def\@cline#1-#2\@nil{%
  \omit
  \@multicnt#1%
  \advance\@multispan\m@ne
  \ifnum\@multicnt=\@ne\@firstofone{&\omit}\fi
  \@multicnt#2%
  \advance\@multicnt-#1%
  \advance\@multispan\@ne
  \leaders\hrule\@height\arrayrulewidth\hfill
  \cr
  \noalign{\vskip-\arrayrulewidth}}
\makeatother

\begin{document}

\title{Instability in Complex Oscillator Networks: Limitations and Potentials of Network Measures and Machine Learning}








\author*[1,2]{\fnm{Christian} \sur{Nauck}}
\equalcont{These authors contributed equally to this work.}
\email{nauck@pik-potsdam.de}

\author[1,3]{\fnm{Michael} \sur{Lindner}}
\equalcont{These authors contributed equally to this work.}

\author[1]{\fnm{Nora} \sur{Molkenthin}}

\author[1]{\fnm{Jürgen} \sur{Kurths}}

\author[3]{\fnm{Eckehard} \sur{Schöll}}

\author[2]{\fnm{Jörg} \sur{Raisch}}

\author*[1]{\fnm{Frank} \sur{Hellmann}}
\email{hellmann@pik-potsdam.de}

\affil*[1]{\orgdiv{RD IV Complexity Science}, \orgname{Potsdam Institute for Climate Impact Research}, \orgaddress{\street{Telegrafenberg A31}, \city{Potsdam}, \postcode{14473}, \state{Brandenburg}, \country{Germany}}}

\affil[2]{\orgdiv{Control Systems}, \orgname{Technical University of Berlin}, \orgaddress{\street{Einsteinufer 17}, \city{Berlin}, \postcode{10587}, \country{Germany}}}

\affil[3]{\orgdiv{Department of Digital Transformation in Energy Systems, Institute of Energy Technology},
\orgname{Technical University of Berlin}, \orgaddress{\street{Einsteinufer 25 (TA 8)}, \postcode{10587}, \country{Germany}}}


\date{\today}

\maketitle
\begin{abstract}

A central question of network science is how functional properties of systems emerge from their structure. For networked dynamical systems, structure is typically captured through network measures. We investigate the relationship between these measures and stability metrics across non-linear and linear oscillators, as well as real-world power grid topologies and dynamics. We find that this relationship is highly sensitive to the underlying ensemble: minor changes in the networks considered, such as going from mean degree 6 to mean degree 8, can invert the correlation between a network measure and stability. We also investigate network measures as inputs for machine learning, as well as Graph Neural Networks (GNNs) as predictors of stability. Both GNNs and the non-linear combination of many network measures can accurately predict stability within a given ensemble, yet both can fail when the ensemble changes. We conclude that neither approach reliably identifies the underlying structural causes of instability.

\end{abstract}
\section*{Introduction}
Networks of coupled oscillators are indispensable for modeling natural and human-made systems. In fields as diverse as biology, neuroscience, ecology, physics, and engineering, important systems including the heart, the brain, food webs, coupled lasers, chemical reactions, power grids, and even firefly populations, are described as oscillators on complex networks \cite{strogatzKuramotoCrawfordExploring2000,acebronSynchronizationPopulationsGlobally2000, pikovskySynchronizationUniversalConcept2001, acebronKuramotoModelSimple2005,pecoraClusterSynchronizationIsolated2014,rodriguesKuramotoModelComplex2016}. The function of these systems is shaped in considerable parts by their connectivity, which is described by the network's topology.

The paradigmatic model used to understand networks of oscillatory systems is the Kuramoto model \cite{kuramotoChemicalOscillationsWaves1984,kuramotoSelfentrainmentPopulationCoupled1975} and its variants. These consist of linear oscillators with a non-linear coupling, and feature extremely rich collective dynamical behavior, such as chimera states, frequency clusters, isolated desynchronization, and spatial chaos \cite{strogatzKuramotoCrawfordExploring2000,acebronSynchronizationPopulationsGlobally2000,acebronKuramotoModelSimple2005, pecoraClusterSynchronizationIsolated2014,rodriguesKuramotoModelComplex2016}. When including amplitude dynamics, one obtains Stuart-Landau oscillators. When combined with specific non-linear couplings, these have been shown to provide realistic models for the dynamics of renewable power grids \cite{koglerNormalFormGridForming2022, buttnerComplexPhaseDataDrivenIdentification2024}. A complementary class of models are given by diffusively coupled non-linear oscillators, with van der Pol oscillators being a prominent example. Whereas the rich behavior and multistability in the Kuramoto setting are driven by the non-linear coupling \cite{hellmannNetworkinducedMultistabilityLossy2020} van der Pol oscillators also feature complex, multistable local dynamics.

The collective phenomena in all these systems are rooted in synchronization \cite{pikovskySynchronizationUniversalConcept2001}. In some contexts, such as the  brain, complete synchronization indicates dysfunction such as epilepsy. In others, such as power grids, it is a fundamental prerequisite for the system to work. Thus, a central question in the fields of networked complex systems is how robust the synchronous state is, either with the aim to design controls to disrupt it, e.g. deep brain stimulation \cite{tassPhaseResettingMedicine2007}, or to create a system that has favorable synchronization properties \cite{yamamotoModularArchitectureFacilitates2023, menckHowBasinStability2013, menckHowDeadEnds2014,bernerWhatAdaptiveNeuronal2021} (for power grids or neuronal networks).

In practice, the structure vs. function relationship is typically studied by considering correlations between network measures, which quantify network structure, and dynamical properties. Two prominent examples of this are i) explosive synchronization \cite{jiClusterExplosiveSynchronization2013}, in which the degree determines the synchronization properties, and ii) the probability that large localized perturbations desynchronize the network or lead to further failures \cite{menckHowDeadEnds2014, hellmannSurvivabilityDeterministicDynamical2016}. The latter is of particular importance in the context of power grids. Localized single component failures occur frequently. If these induce desynchronization of the whole network, large-scale blackouts are the consequence \cite{witthautCollectiveNonlinearDynamics2022}. Understanding the probability that such a desynchronization event occurs turns out to be a particularly challenging task for network and complex systems' science \cite{witthautCollectiveNonlinearDynamics2022}. Despite a decade of effort since Menck~et~al.~\cite{menckHowDeadEnds2014} found that dead ends have unfavorable properties, no clear picture has emerged \cite{kimBuildingBlocksBasin2016,nitzbonDecipheringImprintTopology2017,kimStructuralDynamicalFactors2019}.

Recently, an alternative approach to capture the relationship of structure and function has been developed: graph neural networks (GNNs), which are machine learning (ML) architectures specifically adapted to working with data on graphs. GNNs and their generalizations work well in different domains, such as epidemic spreading via social networks \cite{clipmanDeepLearningSocial2022}, 
and molecular properties \cite{atzGeometricDeepLearning2021}. 
Lately, they also succeeded in predicting the desynchronization probability in oscillator networks \cite{nauckPredictingBasinStability2022,nauckDynamicStabilityAnalysis2022,nauckDynamicStabilityAssessment2023}. The performance reached by these GNNs approaches the level that would make them useful in real-world scenarios. However, it is challenging if not impossible to interpret GNNs and translate the results back to an understanding of the mechanisms in the system. The field of explainability methods for GNNs is still nascent \cite{yuanExplainabilityGraphNeural2022}.

\begin{figure}[ht]
    \centering

    \includegraphics[width=\textwidth]{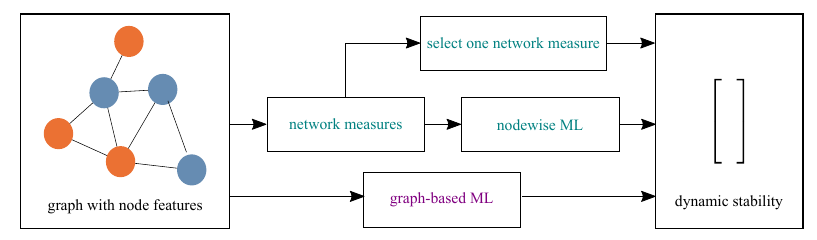}
    \caption{The goal is the prediction of the dynamic stability (targets) based on power grid models (input). Whereas \textcolor{purple_GNN}{GNNs} (at the bottom) deal with the graph input directly, \textcolor{green_NetSciML}{NetSciML} models rely on network measures as inputs.}
    \label{fig:training_setup_simple}
\end{figure}

Here, we curate a wide set of network measures, and systematically explore network science and ML-based approaches to the structure-function relationship. To study non-linear combination of network measures, we make use of node-wise machine learning, that is, we train conventional ML models to predict stability at a node solely based on the network measures associated to that node. This approach to featurizing the nodes using network science will be called NetSciML in the following. In contrast, GNNs consider the whole graph at once, see \Cref{fig:training_setup_simple}. 

We study both transient and asymptotic stability of synchrony in the context of power grid topologies and Watts-Strogatz ensembles, featuring Kuramoto dynamics, Stuart-Landau like power grid models and van der Pol oscillators. 


The central question we address is how robust the relationships identified are. Our central findings are:

\begin{enumerate}
    \item In almost all cases we found that the relationship between network measures and stability properties can vary strongly as parameters of the network ensemble are varied. 
    \item Both GNNs and NetSciML can accurately predict a variety of stability metrics for a range of dynamical networks from the same ensemble. Individual network measures often cannot. Especially in situations featuring heterogeneous dynamical parameters and dynamics, hand-crafted features capturing the dynamical heterogeneity are required to achieve good predictions.
    \item Even if GNNs and NetSciML achieve good predictive power for networks of the same size, drawn from the same ensemble, the predictions of GNNs and NetSciML can fail to generalize to other networks. We observe cases where NetSciML succeeds while GNNs fail and vice versa. We interpret the failure to generalize as a failure to accurately capture the underlying causes of instability.
\end{enumerate}

\section*{Results}

We now show how network measures, and non-linear combinations of network measures with high predictive power, relate to stability of synchrony for very different dynamical systems. We investigate three paradigmatic oscillator models, using appropriate notions of transient and asymptotic stability, on different classes of network structures using a wide range of curated network measures. Our main focus will be how the relationship between network measures and stability metrics changes as the network structure varies.

\textbf{Oscillator models:}  
First, we employ second-order Kuramoto-Sakaguchi oscillators \cite{kuramotoChemicalOscillationsWaves1984,kuramotoSelfentrainmentPopulationCoupled1975,sakaguchiSolubleActiveRotator1986}. These are linear oscillators with non-linear coupling. We use randomly sampled heterogeneous parameters and a more homogeneous power-grid-inspired parameter set. The latter is motivated by the fact that second-order Kuramoto oscillators are a paradigmatic model of the synchronization mechanism in power grids \cite{bergenStructurePreservingModel1981}.

Second, we analyze van der Pol oscillators \cite{cartwbightBalthazarVanPol1960}. These are non-linear oscillators with linear, diffusive coupling. As opposed to the linear oscillators, here multistability and complex behavior is driven by the amplitude multistability of the individual oscillators.

Third, we examine realistic power grid dynamics, focusing on grid-forming inverters relevant for renewable energy integration. We make use of a Stuart-Landau like complex oscillator formulation of grid-forming inverters that has been experimentally validated \cite{koglerNormalFormGridForming2022, buttnerComplexPhaseDataDrivenIdentification2024}. Dynamically they are higher-order linear amplitude-phase oscillators with non-linear coupling. (see \Cref{sec:supplement_inverter_powergrids} for details).

\textbf{Stability metrics:}
We assess both transient and asymptotic stability using probabilistic measures appropriate to the dynamical system \cite{menckHowBasinStability2013,hellmannSurvivabilityDeterministicDynamical2016}. We investigate the stability of the synchronous state to random perturbations occurring at a single node. Thus we have two stability measures per node in each case.

\textbf{Network structure:}
We examine two main classes of network topologies. First, we generate Watts-Strogatz networks \cite{wattsCollectiveDynamicsSmallworld1998} and vary average degree and rewiring probabilities. This way we can systematically explore both, the impact of sparsity, and the transition between regular, small-world, and random network regimes. Second, we analyze power grid topologies spanning a broad range of sizes, primarily generated using a random-growth algorithm that replicates key structural features of real electrical grids \cite{schultzRandomGrowthModel2014}, supplemented by several authentic power grid networks.

\textbf{Experimental Setups}
We study four setups:
\begin{itemize}
    \item \textbf{WS-Kura}: Watts–Strogatz networks with inertial Kuramoto–Sakaguchi dynamics and heterogeneous node and edge parameters.
    \item \textbf{WS-VDP}: Watts–Strogatz networks with van der Pol oscillator dynamics and fully homogeneous parameters.
    \item \textbf{PG-Kura}: Two ensembles of synthetic power-grid like topologies (20 and 100 nodes), the power-grid topologies of four countries (DE, FR, GB, ES) and a large synthetic network of Texas generated using a different synthetic model.
    \item \textbf{PG-Real}: One ensemble of realistic synthetic power-grid topologies (between 70 and 80 nodes) and four larger girds generated the same way (128, 256, 512, 1024 nodes) with high-fidelity inverter-based dynamic models \cite{buttnerComplexPhaseDataDrivenIdentification2024}.
\end{itemize}

\textbf{Network measures:}
We curated a set of established network measures, some of which have been previously used for studying synchronization properties. To account for the various dynamic heterogeneities of the dynamics, we augmented these by network measures derived from the systems parameters as well as properties of the synchronous state: In power grids and Kuramoto oscillators the synchronous state features non-zero phase differences on the edges $\Delta \phi_{ij}$, giving rise to a non-zero "power flow" $\text{Flow}_{ij}$. The system parameters used are the real valued coupling $K_{ij}$, the adjacency matrix $A_{ij}$, the power injection $P^d_i$ and damping $D_i$ for Kuramoto dynamics, and the complex valued $Y_{ij}$ admittance, active and reactive power $P^d$ and $Q^d$ and an abstract "Dynamic type of node" variable that distinguishes different inverters for realistic power grids. The only setup with no extra dynamic network measures is WS-VDP, where all nodes are dynamically identical and the synchronous state has no spatial structure. The full set of network measures used is given in \Cref{sec:app-full-net-measures}.

\textbf{Machine Learning Predictors}
To investigate how informative combinations of network measures can be, we used the node-wise network measures as input for a gradient boosted tree machine learning model, trained to predict the stability metric under investigation (NetSciML). As a further point of comparison, we also trained GNNs to directly predict the stability metrics from dynamic parameters and graph structure (see~\Cref{fig:training_setup_simple}). For WS we chose the average degree 6 and rewiring probability 0.5 as the training ensemble, for PG-Kura we chose the 100 node ensemble and for PG-Real we used the 70-80 node ensemble. We evaluated the predictors in-distribution, on networks drawn from the training ensemble but not used for training, as well as out of distribution on networks structurally different than the training ensemble.

The full set of results is presented in the Supplementary Material (see \Cref{sec:app_Overview}).
    
\subsection*{Pearson correlation of individual network measures and stability varies strongly}

\begin{figure}
    \centering
    \vspace{-.6cm}
    \includegraphics[width=\linewidth]{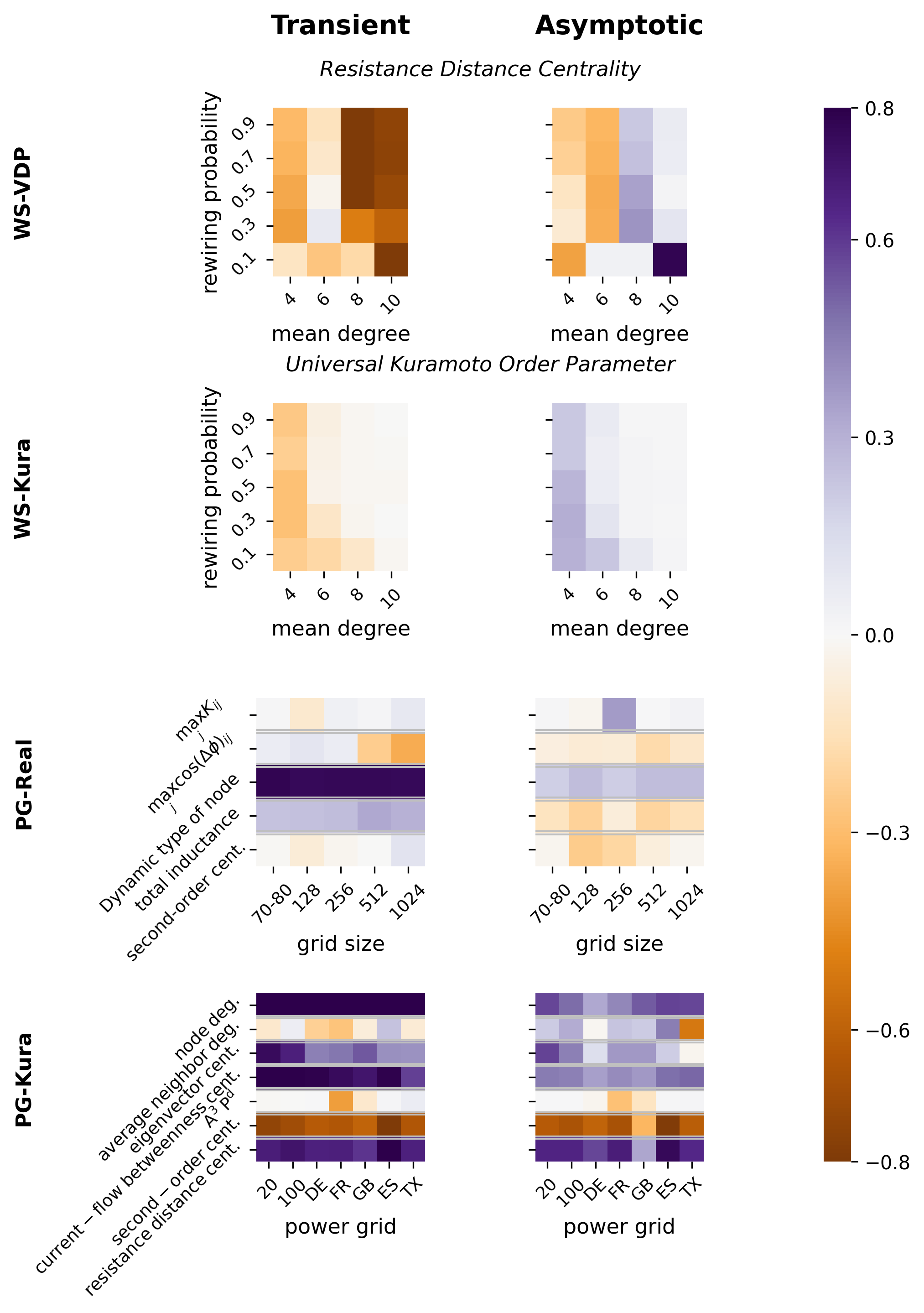}
    \caption{Pearson correlation of network measures across multiple ensembles of different dynamical and topological systems. For WS-VDP, the network measure resistance distance centrality is utilized and the universal Kuramoto order parameter for WS-Kura. For asymptotic stability second order centrality and resistance distance centrality in PG-Kura the values are outside the color band.}
    \label{fig:results_combined_heatmaps_network_measures_pearson_correlation}
\end{figure}

In almost all cases we found that the Pearson correlation between network measures and stability properties varies strongly, and can even flip sign, as parameters of the network ensemble are varied. In \Cref{fig:results_combined_heatmaps_network_measures_pearson_correlation} we present the Pearson correlation between selected network measures and transient and asymptotic stability metrics in all four experiments. To confirm that the Pearson correlation does not miss non-linear relationships we also evaluated Theil's U coefficient. Results for the full set of network measures with both Pearson and Theil's U are provided in the supplemental material (\Cref{sec:app_Overview}). Summary tables of the key results are provided in \Cref{sec:appendix_experiments_summary}; these tables contain most of the detailed numerical values reported below.

In WS-VDP we observe a clear distinction between small and high degree with the boundary between degree 6 and 8, and low and high disorder with the boundary between 0.1 and 0.3 rewiring probability. The correlation of resistance distance centrality and transient stability is $-80\%$ at rewiring probability $0.5$ and mean degree $8$, but $+8\%$ at rewiring probability $0.3$ and mean degree $6$ (see \Cref{tab:VDP pearson selected nm}). In the highly heterogeneous WS-Kura experiment we observe a smoother transition from low to high degrees. The universal Kuramoto order parameter depends on $\Delta \phi_{ij}$, and is correlated to the transient and asymptotic stability at low degrees, but completely uncorrelated at higher degrees. Conversely the dynamical parameters $P^d$ and $D$ show the inverse behavior, they become more informative the higher the degree rises, with $P^d$ showing no Pearson correlation but high Theil's U at higher degree. In PG-Real various measures derived from the synchronous state, such as $\max_j \cos (\Delta \phi_{ij})$ show weakly positive correlation for smaller grids, but become moderately negatively correlated for the larger grids, but there are no clear patterns to the observed changes. In PG-Kura and asymptotic stability we see average neighbor degree going from +0.45 for the Spanish grid to $-0.52$ for the Texan grid (\Cref{tab:PGKura var correlations}). Resistance distance centrality and second order centrality show high correlation across most grids studied, but still vary substantially: from 0.76/-0.79 in the Spanish grid to 0.33/-0.32 for Great Britain (\Cref{tab:PGKura abs correlations}).

Across all experiments, only very few network measures show a significantly non-zero correlation that is stable across topology variation. Those are the dynamic type of the node and the total inductance, which is a weighted degree measure, for PG-Real (\Cref{tab:NF summary}) and current-flow betweenness centrality for PG-Kura asymptotic (\Cref{tab:PGKura summary asymptotic}).


Notably, among the strongly correlated network measures we find both, network measures specifically designed for synchronization dynamics, such as resistance distance centrality \cite{tylooKeyPlayerProblem2019}, as well as network measures that have not previously been considered in this context at all, most importantly second-order centrality\cite{schultzDetoursBasinStability2014,nitzbonDecipheringImprintTopology2017}.

\subsection*{NetSciML and GNNs can predict stability well}
\begin{figure}
    \centering
    \includegraphics[width=\linewidth]{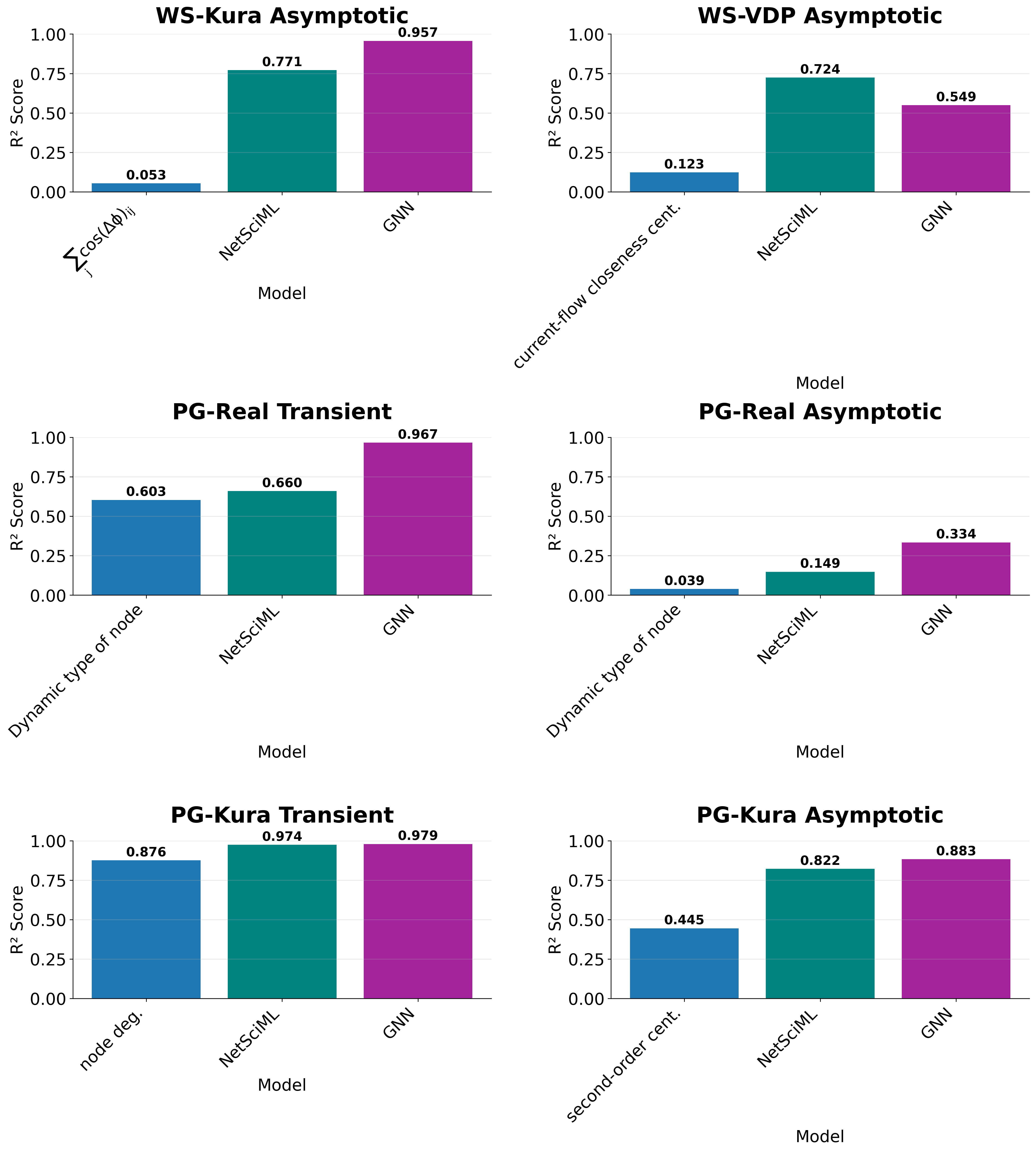}
    \caption{Performance at predicting dynamic stability of the test set of the same ensemble measured by $R^2$. For each task, the single network measure with the highest predictive performance is shown.}
    \label{fig:results_combined_bar}
\end{figure}

\Cref{fig:results_combined_bar} shows the $R^2$ score of linear regression using the best individual network measure, NetSciML and GNNs. The $R^2$ score is evaluated on networks drawn from the same ensemble as the training networks. NetSciML achieves strong predictive performance for all experiments, demonstrating that a non-linear combination of multiple network measures can accurately predict diverse stability metrics across dynamical networks from the same ensemble, almost always significantly improving on the best individual network measure. We also see that GNNs typically outperform NetSciML with the sole exception of predicting asymptotic stability for WS-VDP. There NetSciML achieves 0.72, whereas GNNs only achieve 0.55 (\Cref{tab:VDP NetSciML}).
Especially in WS-Kura, which features heterogeneous dynamical parameters and dynamics, hand-crafted features capturing the dynamical heterogeneity are required to achieve good predictions. Using only topological features, transient predictions reach an $R^2$ of only 0.43 compared with 0.74 when only dynamical features are used. Combining all features yields only a marginal further gain, increasing $R^2$ to 0.75 (\Cref{tab:WSKura summary}).

\begin{figure}
    \centering
        \includegraphics[width=.7\linewidth]{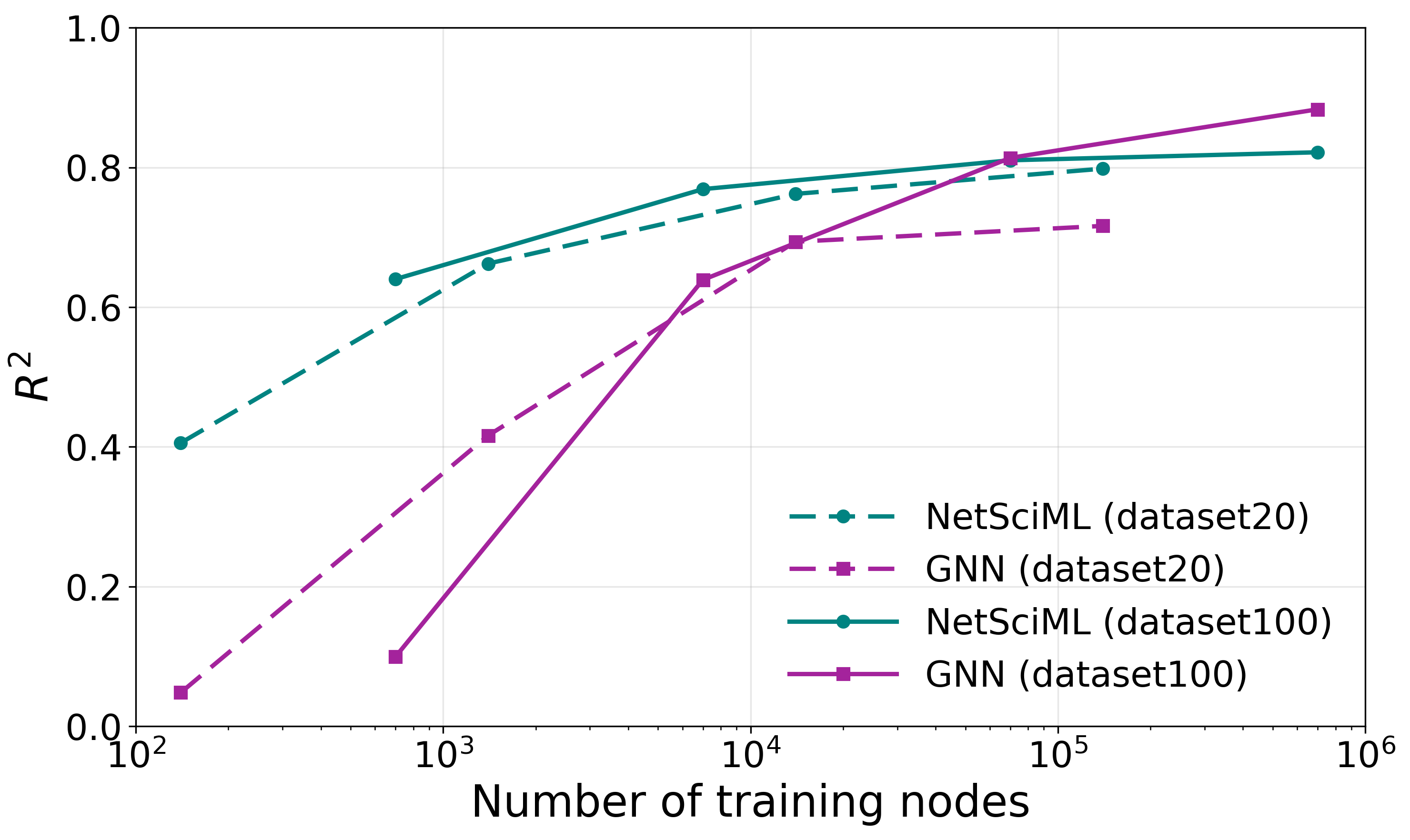}
    \caption{Performance on asymptotic stability in PG-Kura, measured by $R^2$, as a function of the amount of available training data. The plot compares GNNs and NetSciML on dataset20 (dashed) and dataset100 (solid).}
    \label{fig:more_data_GNN_vs_NetSciML}
\end{figure}

As \Cref{fig:more_data_GNN_vs_NetSciML} shows for predicting asymptotic stability for PG-Kura, NetSciML can achieve its performance with considerably less training data than required by GNNs. When comparing the performances on both ensembles consisting of grids of size 20 or 100, the results suggest that the achievable performance depends more on the total number of nodes in the training set, than on the total number of grids or the grid size. This demonstrates clearly the utility of combining theory and data driven analysis in situations where data is comparatively sparse.

\subsection*{GNNs and NetSciML can fail to capture the root causes of instability}

\begin{figure}[h]
    \centering
    \includegraphics[width=\linewidth]{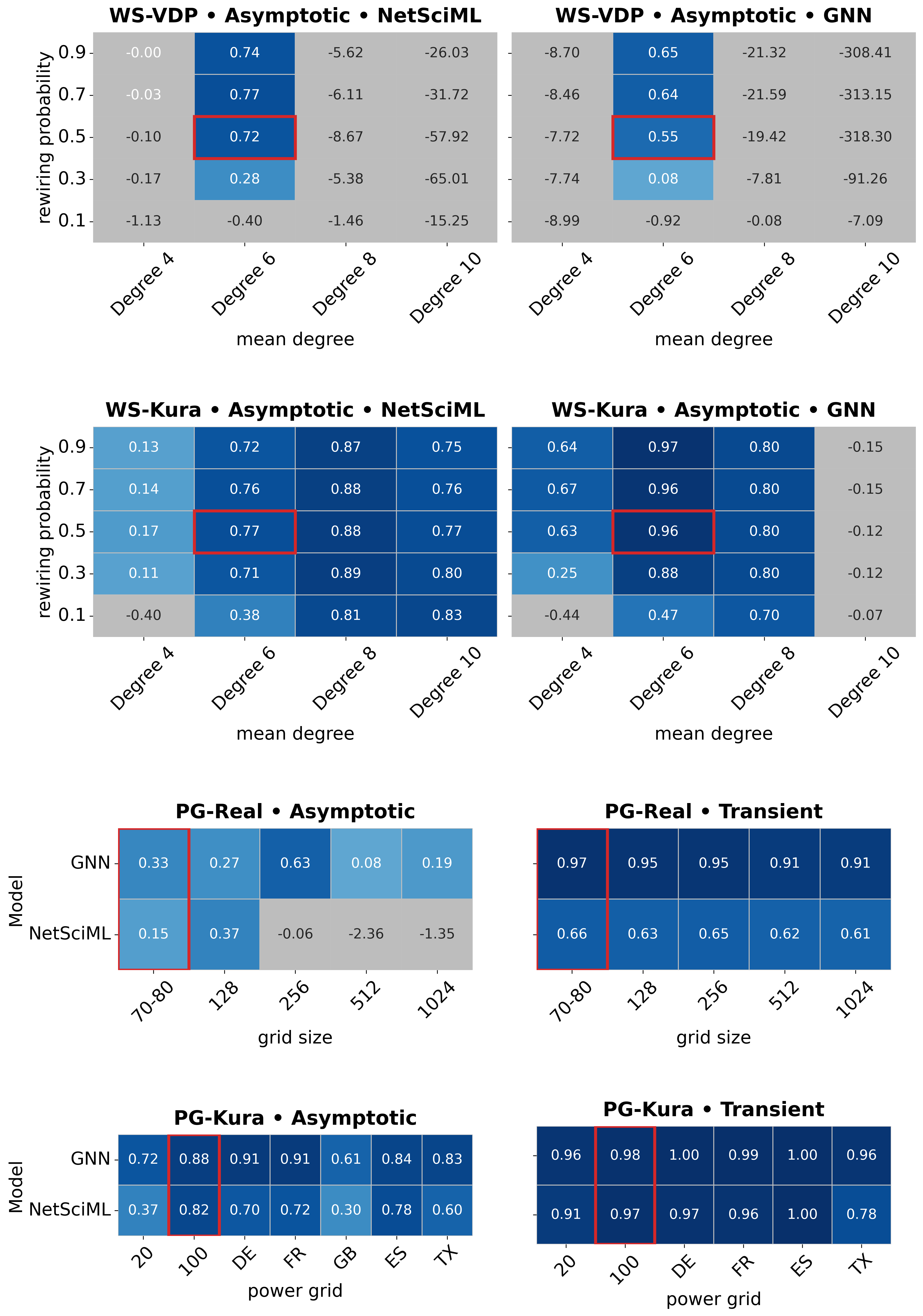}
    \caption{Performance at predicting dynamic stability across different ensembles measured by $R^2$. The red rectangle mark the training ensembles.}
    \label{fig:results_combined_netsciml_vs_gnn}
\end{figure}

Even though NetSciML and GNNs achieve good performance when evaluated on networks drawn from the same ensemble, they can fail if evaluated on networks from other ensembles, indicating that they learn spurious correlations rather than root causes of instability.
We evaluated the performance of the trained predictors for asymptotic stability for all experiments, and for transient stability for the PG experiments on all variations of the networks. Results are given in \Cref{fig:results_combined_netsciml_vs_gnn}.

For power grid tasks (PG-Kura and PG-Real), GNNs achieve higher predictive performance than NetSciML; both tend to lose performance when predicting on other ensembles or real-world topologies, but GNNs appear considerably more robust. For PG-Kura asymptotic stability, within ensemble, GNNs and NetSciML achieve 0.88 and 0.82 $R^2$. When evaluated on synthetic grids of size 20, these drop to 0.72 for and GNN but 0.38 for NetSciML. For PG-Kura transient stability, we see only a minor drop from 0.98 to 0.96 for GNNs when evaluating on the large Texas power grid, but NetSciML drops from 0.97 to 0.78.

For PG-Real, transient performance is flat and generalizes perfectly, indicating that in this case the predictors are actually capturing the underlying causes of instability. For asymptotic stability, GNNs achieve positive $R^2$ across all grid sizes, while NetSciML does so only to size 128 and fails at larger scales.

For WS-VDP, neither NetSciML nor GNNs generalizes across ensembles with different average degrees, suggesting that neither method captures causal relations that transfer across regimes. Nonetheless, NetSciML retains its in-distribution advantage over GNNs for the same average degree. The shifts in network-measure correlations across ensembles already discussed, indicate substantial changes in dynamical behavior, which likely explain the difficulty both models face in learning transferable patterns.

For WS-Kura, NetSciML generalizes much better to high degree networks than GNNs, despite GNNs achieving much stronger in-distribution performance. At average degree 10, GNNs fail to predict dynamical stability, whereas NetSciML maintains strong predictive accuracy. Notably, NetSciML even performs better in some cross-ensemble settings than within ensemble. Across average degrees, several network measures exhibit clear structure, with strong correlations concentrated in either low- or high-degree regimes. The predictive results suggest that NetSciML correctly identifies dynamical patterns associated with larger average degrees; as these patterns become more dominant with increasing degree, predictive performance correspondingly improves.

It is also notable that NetSciML predictions of asymptotic stability using only topological features exhibit a similar pattern as GNNs here (\Cref{tab:WSKura summary}), with a sharp reduction of predictive performance when going from degree 8 to 10. This indicates that GNN accurately identify topological causes for instability but miss the dynamic patterns that become dominant at larger degrees. This also aligns with the fact that GNNs generalize better to degree 4 networks.

As expected, the ability of NetSciML to generalize robustly corresponds to stable correlation patterns of the underlying network measures. For example for transient stability in power grids, node degree predicts stability across ensembles for PG-Kura (\Cref{tab:PGKura summary asymptotic,tab:PGKura summary transient}) and, and node type for PG-Real: (\Cref{tab:NF summary}). If the explanatory power of individual network measures varies dramatically, the network measure based non-linear predictors struggle to generalize.

\clearpage
\newpage

\FloatBarrier
\section*{Discussion}

Much of the network theoretic analysis of asymptotic stability of power grid synchronization of the past decade implicitly assumed that network measures could capture the underlying causes for high or low basin stability, e.g., that dead ends undermine stability \cite{menckHowDeadEnds2014}. However, if network measures were causally determinative for stability, one would expect this relationship to be robust. 

Instead we see that the relationship of stability metrics and network structure as captured by network measures is highly sensitive to the specific network ensemble the network comes from. Even relatively modest changes in the network ensemble, such as going from mean degree 6 to 8 to 10 can render observed correlations irrelevant or even completely misleading.

Similar observations hold for stability predictions using NetSciML or GNNs. Both can fail to predict stability in networks that are not dramatically different from the ones they are trained on. Correlations between network measures and stability, as well as GNN and NetSciML predictions should not be interpreted causally without careful validation. This poses a significant challenge for learning about dynamic phenomena from the analysis of such predictors using explainability methods.

For all these challenges, it is also notable that it always was possible to obtain good predictions of dynamic stability from structural information within the ensemble by drawing on a broad range of network measures. Using either method, we obtained $R^2 > 0.5$ everywhere except for PG-Real Asymptotic. As expected, NetSciML was considerably more data efficient. We also observed that handcrafted features capturing heterogeneity of the synchronous state and the nodal dynamics are crucial to obtaining good performance in the NetSciML approach.

It remains unclear what determines the sensitivity of the network measure - stability relationship. WS-VDP is the most homogeneous experiment we ran, and showed the strongest sensitivity. It is plausible that dynamic heterogeneity attenuates the sensitivity. On the other hand we observed catastrophic failure to generalize in GNN predictions for WS-Kura, our most heterogeneous experiment.

An open question out of scope for this work is whether these sensitivities are typical more broadly for structure - function type relationships. It would be highly interesting to study dynamical properties other than stability, as well as dynamic networks that arise in other real-world contexts, especially neuronal dynamics and molecular dynamics. In the latter context, GNNs are already widely used. Hand-crafting features for nodes has also been explored, but to our knowledge not by using the systematic network-science based approach we introduce here. A central challenge remains generating appropriate datasets.

A further challenge lies in hand-crafting appropriate network measures to capture dynamic heterogeneity. No principled way to distill heterogeneous dynamic properties into network measures, or to select topological measures particularly pertinent to certain dynamic settings, exist. The fact that one of the most informative network measures for the much studied PG-Kura setting, second order centrality, was only serendipitously included here highlights how challenging this task is.

For the PG-Kura setting, the possibility to combine network measures and GNNs was explored in \cite{zhuNetworkMeasureEnrichedGNNs2026}. There it was shown that naively using the network measures as additional GNN input features is not a useful strategy unless it is known that there are no major distribution shifts between network measures. Instead a more robust approach was to use the GNN to predict both, stability and network measures, forcing the GNN to more robustly characterize the underlying graph.

From a practical perspective, the data efficiency of the NetSciML approach is promising for enabling predictions in data-sparse applications such as power grids. Central challenges remain the need for hand-crafted features and the lack of robust generalization. It remains an open question whether combinations of GNNs and NetSciML could also improve the data efficiency of GNNs. 

Our results do not allow drawing strong conclusions on why NetSciML and GNNs sometimes fail to generalize for modest structural variations. Two essential possibilities are that they learn spurious correlations that change as the ensemble changes, or that the mechanism of instability itself actually changes. In the former case, no (direct) causal conclusions can be drawn from studying the predictors. In the latter, a careful analysis of the within ensemble predictions would still be causally meaningful. The results for WS-Kura indicate that in this case GNNs suffer from a failure of the first kind, but for the other experiments our results do not provide
evidence either way. Thus, the central theoretical question raised by our findings is  why we see such dramatic failures and sensitivity.

\FloatBarrier
\section*{Methods}

\subsection*{Networks and Network Ensembles}
We consider two main classes of networks: (i) Watts-Strogatz networks and (ii) power grids with both synthetic and real-world topologies.

\subsubsection*{Watts-Strogatz networks}
For the Watts-Strogatz networks \cite{wattsCollectiveDynamicsSmallworld1998}, we systematically vary the mean degree and rewiring probability to generate ensembles with diverse structural properties. We consider four average degrees (4, 6, 8, and 10) and five rewiring probabilities (0.1, 0.3, 0.5, 0.7, 0.9).

For the Kuramoto oscillator experiments, we generate 2,048 networks per ensemble, except for rewiring probability of 0.5 and degree 6, where we generate 20,480 networks to provide larger training sets.

For the van-der-Pol oscillator experiments, 1,024 networks per ensemble are generated, except for rewiring probability of 0.5 and degree 6, where 21,504 networks are generated.

\subsubsection*{Ensembles of synthetic and real world power grid models}

To systematically investigate the relationship between network measures and dynamic stability in power grid models, we utilize large datasets from \cite{nauckDynamicStabilityAnalysis2022,nauckDynamicStabilityAssessment2023}. These datasets, specifically designed for machine learning applications, comprise 10,000 networks each for ensembles with 20 and 100 nodes, as well as a synthetic model of the Texan power grid with 1,910 nodes (TX).

To assess the generalizability of our ML models from synthetic to real-world topologies, we further analyze four high-voltage power grid models based on the actual network structures of France (FR), Great Britain (GB), Germany (DE), and Spain (ES). The French (146 nodes), British (120 nodes), and Spanish (98 nodes) topologies are sourced from \cite{schaferDynamicallyInducedCascading2018}, while the German grid (438 nodes) is provided by the German Institute for Economic Research (DIW) \cite{egererOpenSourceElectricity2016}. For all real-world grids, we parameterize nodes and edges following the procedure established for the Texan grid in \cite{nauckDynamicStabilityAnalysis2022,nauckDynamicStabilityAssessment2023}.

\subsubsection*{Realistic power grids}
To further validate the robustness of our methods, we apply them to a more realistic power grid dataset introduced in \cite{nauckPredictingFaultRideThroughProbability2026}. This dataset features advanced node modeling, with loads represented by algebraic constraints and producers parameterized as inverters in three configurations. Line modeling incorporates losses, and admittances are based on the real-world German power grid. The synthetic grid generation framework is described in \cite{buttnerFrameworkSyntheticPower2023}, which includes validation steps to ensure realistic topological and dynamical properties. The original dataset contains 1,000 grid samples for training, each with 70–80 nodes. To test generalization across grid sizes, we also conduct simulations on larger grids: G128 (128 nodes), G256 (256 nodes), G512 (512 nodes), and G1024 (1,024 nodes).

\subsection*{Dynamical models}

We investigate three types of oscillator dynamics: (i) the second-order Kuramoto-Sakaguchi model, (ii) van der Pol oscillators, and (iii) inverter dynamics relevant for realistic power grids with high shares of renewable energy sources.

\textbf{Second-order Kuramoto-Sakaguchi model:}  
The paradigmatic second-order Kuramoto model \cite{kuramotoSelfentrainmentPopulationCoupled1975} is given by
\begin{equation}
    \ddot{\phi}_i = P^d_i - D_i \dot{\phi}_i - \sum_{j=1}^n K_{ij} A_{ij} \sin(\phi_i - \phi_j - \alpha_{ij}),
    \label{eqKuramoto}
\end{equation}
where $\phi_i$ is the phase angle at node $i$, and $\dot{\phi}_i$, $\ddot{\phi}_i$ are its first and second time derivatives, respectively. The network topology is encoded in the adjacency matrix $A_{ij}$.

In PG-Kura we use a homogeneous bimodal parametrization that is meant to capture conceptual properties of future transmission grids with distributed resources: $P^d_i \in \{-1, 1\}$ indicates net-consumer or producer and is randomly drawn, damping $D_i = 0.1$, and a relatively large, homogeneous overall coupling strength $K_{ij} = 9$ and $\alpha_{ij} = 0$. A synchronous state typically exists for this system; however, it is often not globally stable. For WS-Kura all parameters are drawn uniformly at random for every line/node, with $\alpha_{ij} \in [0, 1]$, $K_{ij} \in [0, 4]$, $D_i \in [-0.15, -0.05]$ and $P^d_i$ from a normal distribution of width $1$ and centered at $0$.

\textbf{Van der Pol oscillator:}  
The van der Pol oscillator is a classical nonlinear oscillator that exhibits both amplitude and phase dynamics. The equations for a network of coupled van der Pol oscillators are:
\begin{align}
    \ddot{x}_i = - x_i + \mu (1 - x_i^2 + \alpha x_i^4 - \beta x_i^6) \dot{x}_i - \sum_{j=1}^n A_{ij} (x_i - x_j) \ ,
    \label{eqVdP}
\end{align}
where $x_i$ is the state at node $i$. We considered a homogeneous population with $\mu = 0.01$, $\alpha = 0.1$ and $\beta = 0.002$, a dynamically rich regime studied in \cite{ngansoTwoattractorChimeraSolitary2023}.

\textbf{Inverter dynamics:}  
For realistic power grid modeling, we use inverter-based dynamics, which are crucial for grids with high shares of renewable energy. The  modeling is based on the normal form, introduced by \cite{koglerNormalFormGridForming2022} and further validated by \cite{buttnerComplexPhaseDataDrivenIdentification2024}. Importantly this model does not only consider frequency, but also voltage dynamics and are generalized Stuart-Landau equations. We have an oscillator $v_i \in \mathbb{C}$ and scalar $x_i$ at each node, the coupling is given in terms of a complex admittance matrix $Y_{ij}$.

\begin{align}
S_{i} &= v_i \sum_j \overline Y_{ij} \overline v_j - P^d_i - j Q^d_i\\
\frac{\dot v}v &= \mu_1 + \mu_2 |v|^2 v + \mu_3 S_i + \mu_4 \overline S_i + \mu_5 x\\
\dot x &= \mu'_1 + \mu'_2 |v|^2 v + \mu'_3 S_i + \mu'_4 \overline S_i + \mu'_5 x\; .
\end{align}

$P^d$ and $Q^d$ are the real and reactive power injected at a node, $S_i$ is the complex power imbalance. Passive load nodes are modeled by a constraint $S_i = 0$ enforcing a constant active and reactive power draw. Following \cite{nauckPredictingFaultRideThroughProbability2026} we have three types of inverters corresponding to three different sets of coefficients $\mu$.

\subsection*{Stability metrics}

To ensure robust and meaningful results, we employ probabilistic stability metrics, considering both transient and asymptotic stability. The specific metric used depends on the task and system under investigation.

\subsubsection*{Asymptotic stability}

\begin{figure}
   \centering
   \includegraphics[width=.7\linewidth]{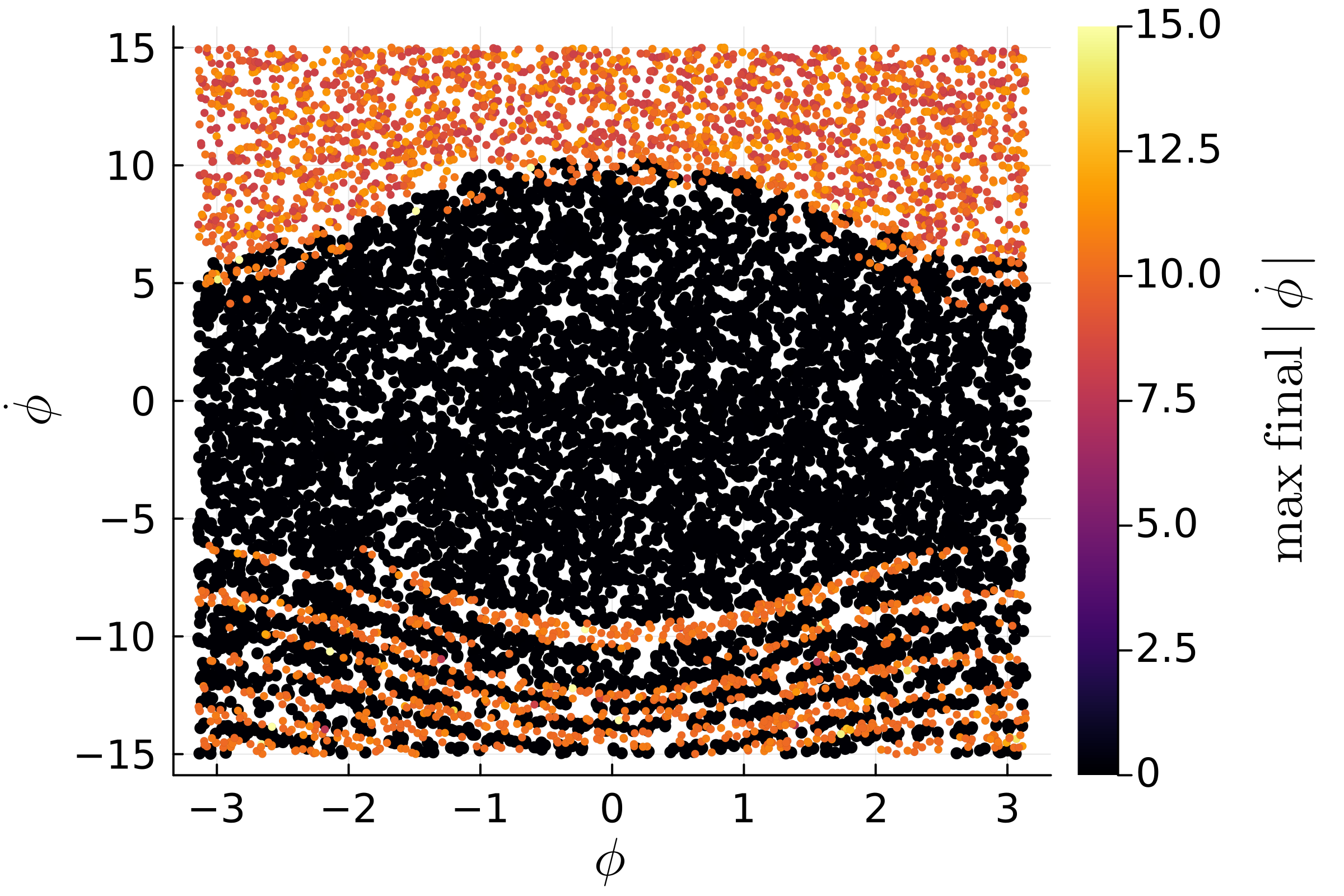}
   \caption{Basin landscape of a node of a 20-node grid with relatively low stability ($\mathrm{SNBS} \approx 0.67$). The color indicates the maximum absolute frequency deviation of all nodes at the end of the trajectory. Black initial conditions converge back to the synchronous state, others reach desynchronized states. SNBS is equal to the fraction of black points among all 10,000 perturbations.}
   \label{fig:basin_snapshots}
\end{figure}

The asymptotic stability measure we use is the single-node basin stability (SNBS) metric \cite{menckHowDeadEnds2014}. SNBS quantifies the probability that, after a large random perturbation applied to a single node, the system returns to synchrony. Mathematically, SNBS corresponds to the volume (with respect to the perturbation probability measure) of the basin of attraction of the synchronous state, restricted to perturbations at a single node. An example of such a phase space slice for PG-Kura is shown in \Cref{fig:basin_snapshots} for a node with relatively low stability.

For inverter-based power grids, we extend the SNBS concept to account for both frequency and voltage stability. A configuration is only considered stable if it satisfies both criteria following a perturbation.

\subsubsection*{Transient stability}
For Kuramoto power grids, we use the maximum frequency deviation metric from  \cite{nauckDynamicStabilityAssessment2023}, defined as the largest frequency deviation during the transient period.

In the context of realistic power grids, we use the probabilistic ride-through probability ($\pfrt$) as introduced in \cite{nauckPredictingFaultRideThroughProbability2026}, a variation of the survivability \cite{hellmannSurvivabilityDeterministicDynamical2016}. This metric measures the likelihood that a node or system does nto violate operational boundaries (i.e., “rides through”) after a disturbance such as a fault or sudden operational change. The stability limits are time-dependent, becoming stricter as time progresses to ensure the system returns to a stable state within a predefined window. For these models, both frequency and voltage stability are evaluated simultaneously.

For van der Pol oscillator networks, we use the \textit{Synchronization Norm} considered in \cite{tylooRobustnessSynchronyComplex2018,tylooKeyPlayerProblem2019}, the time averaged variance of the state $x_i(t)$.

\subsection*{Informativity measures}

To quantify the relationship between network measures and dynamic stability, and the performance of predictors, we employ several informativity metrics: Pearson correlation, Theil's U, and the coefficient of determination ($R^2$).

\textbf{Pearson correlation} ($r$) measures the linear correlation between two variables $X$ and $Y$. It is defined as
\begin{equation}
    r = \frac{\mathrm{cov}(X, Y)}{\sigma_X \sigma_Y},
\end{equation}
where $\mathrm{cov}(X, Y)$ is the covariance and $\sigma_X$, $\sigma_Y$ are the standard deviations of $X$ and $Y$, respectively. $r$ ranges from $-1$ (perfect negative linear correlation) to $1$ (perfect positive linear correlation), with $0$ indicating no linear correlation.

\textbf{Theil's U} is a normalized measure of mutual information between random variables. It is defined as the fraction of the total information describing of $X$ we can predict using $Y$:
\begin{equation}
    U(Y|X) = \frac{H(Y) - H(Y|X)}{H(Y)},
\end{equation}
where $H(Y)$ is the entropy of $Y$ and $H(Y|X)$ is the conditional entropy of $Y$ given $X$. Theil's U ranges from $0$ (no information) to $1$ (perfect information). We estimated U by binning stability and network measures into 25 bins.

\textbf{Coefficient of determination} ($R^2$) is used to assess the performance of NetSciML, GNNs and to compare the predictive power of network measures. It is given in terms of the mean square error $\mse$:
\begin{equation}
    R^2 := 1 - \frac{\mse(\widehat y,y)}{\mse(y_\mathrm{mean},y)},
    \label{eq:R2}
\end{equation}
where $\hat{y}$ are the predictions, $y$ are the target values, and $y_{\mathrm{mean}}$ is the mean of the target values in the test set. $R^2$ measures the proportion of variance in the data explained by the model, with $R^2 = 0$ corresponding to a model that predicts the mean. Negative $R^2$ indicates a performance worse than predicting the mean, and thus a complete misprediction.

Together, these informativity measures provide a comprehensive assessment of the relationships between network measures, predictors and dynamic stability, capturing both linear and nonlinear dependencies as well as model performance.

\subsection*{Network measures}

A table listing the complete set of network measures and the models to which they apply is provided in \Cref{sec:app-full-net-measures}. The table also reports each measure’s origin, and which experimental setting they apply to. Because many measures appear in multiple sources, we list at most two references per measure.

To capture the structural properties of the networks in this study, we assembled a comprehensive set of network measures informed by an extensive literature review. This collection includes measures previously proposed as relevant for oscillator stability, as well as additional features developed in the broader context of network science. We further expanded this set with custom measures and additional metrics available in the NetworkX software package \cite{hagbergExploringNetworkStructure2008}.

One common application of oscillator stability analysis is power grids, so we place strong emphasis on identifying all relevant network measures from this domain. In one of the earliest applications of probabilistic dynamic stability assessment of power grids \cite{menckHowDeadEnds2014} found that leaf nodes correlate with low stability. Subsequently, \cite{schultzDetoursBasinStability2014} predicted single-node basin stability (SNBS) of power grids with logistic regression. They classified nodes as belonging to certain network motifs, such as tree-like subgraphs, and used these as inputs for the regression model, alongside the injected power $P^d$ and the common network measures, degree, average neighbor degree, clustering coefficient, current-flow betweenness and closeness centrality.

\cite{nitzbonDecipheringImprintTopology2017} expand upon this work by studying survivability as well as SNBS, and by refining the topological classification scheme. To compute the proposed node categories \cite{nitzbonDecipheringImprintTopology2017}, the open source software from GitHub \cite{schultzLuappiktreenodeclassification2021} was used. Furthermore, we consider connections between nodes of walks of lengths higher than 1, by computing the row-sum of $A^k$, where $k=3$ would represent walks of length 3.

To identify local vulnerabilities in power grids with intermittent fluctuations, \cite{tylooRobustnessSynchronyComplex2018,tylooKeyPlayerProblem2019} use centrality measures based on resistance distance and Kirchhoff indices. Feld et al.~\cite{feldLargedeviationsBasinStability2019} construct very stable and very unstable power grids and study the correlation of basin stability with various power grid specific network measures, such as flow backup capacity, power sign ratio \cite{dewenterLargedeviationPropertiesResilience2015} and  universal Kuramoto order parameter~\cite{schroderUniversalOrderParameter2017}. 
\cite{yangPowergridStabilityPredictions2021} predict re-synchronization after individual perturbations in synthetic grid topologies generated with the same model that we use~\cite{schultzRandomGrowthModel2014}. As input features for their ML models, they use a few common network measures, such as eigenvector centrality and $k$-core index. Furthermore, they use the specific perturbation vector, and the so-called community inconsistency. The \emph{community inconsistency} depends on a free parameter for which the authors do not provide a heuristic choice \cite{kimCommunityConsistencyDetermines2015}. Since, furthermore, \cite{yangPowergridStabilityPredictions2021} find no large correlation of this measure with dynamic stability, we do not include it in our study.

The related problem of predicting critical links in power grids has been studied by Witthaut et al.~\cite{witthautCriticalLinksNonlocal2016}. For this problem, Titz et al.~\cite{titzPredictingDynamicStability2024} used Gradient Boosted Trees to predict de-synchronization event after line failures at high accuracy. They include many network measures as features, however most of them are only applicable to lines.  

To capture the heterogeneity of the dynamics and the synchronous state we further added hand-crafted measures where appropriate. We considered the dynamic parameters $P^d$, $Q^d$, $D$, where they apply, and a categorical node type feature for PG-Real. For PG-Real, PG-Kura and WS-Kura, the synchronous state is determined by phase angles $\phi$. To capture the local properties of the synchronous state we calculated edge wise $\Delta \phi_{ij}$, $\cos(\Delta \phi_{ij})$ and the total flow on the edge $\text{Flow}_{ij} = K_{ij} \sin(\Delta \phi_{ij} - \alpha_{ij})$. We then constructed nodal measures by taking the sum, minimum, maximum, and median of these edge wise values for each edge incident on a node. For PG-Real, the flow is calculated as the magnitude of the average apparent power on the line, and we further include various measures that characterize $Y_{ij}$, including the total and shunt inducance.

\subsection*{Dynamic stability from structure - NetSciML}

We employ two supervised machine learning approaches to predict dynamic stability labels: graph neural networks (GNNs) and NetSciML. The GNN-based setup is described in the following section; here, we focus on the NetSciML approach.

To systematically assess the predictive power of network measures for capturing the structure–function relationship, we use gradient boosted trees to predict stability from the full set of network measures. Histogram Gradient Boosted Regressors were used for all predictions using the library scikit-learn \cite{PedregoseScikitPython2011}.

\subsection*{Dynamic stability from structure - GNNs}
Graph neural networks (GNNs) are artificial neural networks designed for learning  relationships within data structured as graphs. Their adjustable internal weights can be optimized to suit the specific task at hand. They take the graph structure itself as input, along with optional features associated to the nodes and edges. GNNs can generate outputs in the form of global attributes that describe the entire graph, attributes specific to sub-graphs, or properties associated with individual nodes or edges.

To capture long-range dependencies—an important property for analyzing our problem—we focus on two GNN architectures: (a) Topology Adaptive Graph Convolution (TAGNet) [\cite{duTopologyAdaptiveGraph2017}], which incorporates information from small subgraphs and indirect neighbors within a single layer, and (b) Dirac–Bianconi graph neural networks (DBGNNs) [\cite{nauckDiracBianconiGraph2024}].
In contrast to conventional GNNs, which propagate information diffusively (analogous to the heat equation), DBGNNs enable coherent long-range feature propagation. This mechanism has shown clear benefits for prediction tasks on power-grid networks.

The GNN layers used in this study are described in detail in \Cref{appendix_GNN} and we rely on PyTorch Geometric \cite{feyFastGraphRepresentation2019}.

\FloatBarrier


\section*{Acknowledgments}
All authors gratefully acknowledge Land Brandenburg for supporting this project by providing resources on the high-performance computer system at the Potsdam Institute for Climate Impact Research. Michael Lindner greatly acknowledges support by the Berlin International Graduate School in Model and Simulation based Resarch (BIMoS)  and Christian Nauck would like to thank the German Federal Environmental Foundation (DBU) for funding his PhD scholarship. Further, the work was supported by the Deutsche Forschungsgemeinschaft (DFG, German Research Foundation, HE 6698/4-1 and KU 837/39-2) and the German Federal Ministry for Economy and Energy (03EI1092A). Special thanks to Anna Reckwitz for sharing her code for computing the semi-analytic bound for survivability. AI Tools such as LanguageTool and ChatGPT are used to improve the grammar and wording on a sub-sentence level. We also want to thank Konstantin Schürholt for carefully reading the manuscript and his valuable feedback.

\section*{Data Availability Statement}
\label{sec:data_availability}
The used datasets of the ensembles and the synthetic Texan power grid are available at: \url{https://zenodo.org/record/6572973} and we will also publish all code for the training of the ML models (NetSciML and GNNs), the semi-analytic boundary, the results of the real power grid topologies, as well as code to generate the figures upon publication on Zenodo: \url{https://doi.org/10.5281/zenodo.10686691} and Github: \url{https://github.com/PIK-ICoNe/NetworkScienceML_paper-companion}. It will be licensed under the Creative Commons Attribution 4.0 International license (CC-BY 4.0).

\newpage

\appendix
\FloatBarrier
\renewcommand{\thesection}{SM\arabic{section}}
\renewcommand{\thetable}{ST\arabic{table}}
\crefalias{table}{apptable}

\section{Supplementary material}
\label{sec:app_Overview}

The supplementary material is structured as follows. \Cref{sec:app-full-net-measures} shows the full set of network measures. \Cref{sec:appendix_experiments_summary} contains a summary of the experimental results, followed by detailed results for each experiment (\Cref{sec:appendix_experiments_WSVDP,sec:appendix_experiments_WSKura,sec:appendix_experiments_PGKura,sec:appendix_experiments_PGReal}). \Cref{sec:app_MLMethods} contains supplementary material on the ML methods.

\section{The full set of network measures}
\label{sec:app-full-net-measures}
\Cref{tab:Network measures} lists all network measures used, along with details on their origin and application in this study.

\begin{longtable}{p{7cm}|r|ccc}
\caption{List of network measures used. If the network measure is applicable in the ensemble, $\checkmark$ indicates that it varies per node, $\square$ indicates that it varies per network.} \label{tab:Network measures} \\
\toprule
Network Measure & Origin & Kuramoto & Real. PG & VDP \\
\midrule
\endfirsthead
\caption[]{\textbf{(continued)} List of network measures used. If the network measure is applicable in the ensemble, $\checkmark$ indicates that it varies per node, $\square$ indicates that it varies per network.} \\
\toprule
Network Measure & Origin & Kuramoto & Real. PG & VDP \\
\midrule
\endhead
\midrule
\multicolumn{5}{r}{Continued on next page} \\
\midrule
\endfoot
\bottomrule
\endlastfoot
\textsf{average neighbor deg.} & \cite{schultzDetoursBasinStability2014} & $\checkmark$ & $\checkmark$ & $\checkmark$ \\
\textsf{betweenness cent.} & \cite{yangPowergridStabilityPredictions2021} & $\checkmark$ & $\checkmark$ & $\checkmark$ \\
\textsf{closeness cent.} & \cite{schultzDetoursBasinStability2014} & $\checkmark$ & $\checkmark$ & $\checkmark$ \\
\textsf{clustering coefficient} & \cite{schultzDetoursBasinStability2014, feldLargedeviationsBasinStability2019} & $\checkmark$ & $\checkmark$ & $\checkmark$ \\
\textsf{current-flow betweenness cent.} & \cite{schultzDetoursBasinStability2014} & $\checkmark$ & $\checkmark$ & $\checkmark$ \\
\textsf{current-flow closeness cent.} & \cite{tylooKeyPlayerProblem2019} & $\checkmark$ & $\checkmark$ & $\checkmark$ \\
\textsf{node deg.} & \cite{schultzDetoursBasinStability2014, yangPowergridStabilityPredictions2021} & $\checkmark$ & $\checkmark$ & $\checkmark$ \\
\textsf{eccentricity} & (networkx) & $\checkmark$ & $\checkmark$ & $\checkmark$ \\
\textsf{eigenvector cent.} & \cite{yangPowergridStabilityPredictions2021} & $\checkmark$ & $\checkmark$ & $\checkmark$ \\
\textsf{Fiedler eigenvector} & \cite{wangNetworkApproachPower2015, witthautCriticalLinksNonlocal2016} & $\checkmark$ & $\checkmark$ & $\checkmark$ \\
\textsf{harmonic cent.} & (networkx) & $\checkmark$ & $\checkmark$ & $\checkmark$ \\
\textsf{Katz cent.} & (networkx) & $\checkmark$ & $\checkmark$ & $\checkmark$ \\
\textsf{load cent.} & (networkx) & $\checkmark$ & $\checkmark$ & $\checkmark$ \\
\textsf{maximum neighbor deg.} & (ours) & $\checkmark$ & $\checkmark$ & $\checkmark$ \\
\textsf{minimum neighbor deg.} & (ours) & $\checkmark$ & $\checkmark$ & $\checkmark$ \\
\textsf{resistance distance cent.} & \cite{tylooKeyPlayerProblem2019} & $\checkmark$ & $\checkmark$ & $\checkmark$ \\
\textsf{row-sum $A^2$} & (ours) & $\checkmark$ & $\checkmark$ & $\checkmark$ \\
\textsf{row-sum $A^3$} & (ours) & $\checkmark$ & $\checkmark$ & $\checkmark$ \\
\textsf{second-order cent.} & (networkx) & $\checkmark$ & $\checkmark$ & $\checkmark$ \\
\textsf{square clustering coefficient} & (networkx) & $\checkmark$ & $\checkmark$ & $\checkmark$ \\
\textsf{$A^3 P^d$} & (ours) & $\checkmark$ & $\checkmark$ &  \\
\textsf{$A^2 P^d$} & (ours) & $\checkmark$ & $\checkmark$ &  \\
\textsf{$A P^d$} & (ours) & $\checkmark$ & $\checkmark$ &  \\
\textsf{$\max_j K_{ij}$} & (ours) & $\checkmark$ & $\checkmark$ &  \\
\textsf{$\median_j K_{ij}$} & (ours) & $\checkmark$ & $\checkmark$ &  \\
\textsf{$\min_j K_{ij}$} & (ours) & $\checkmark$ & $\checkmark$ &  \\
\textsf{$\sum_j K_{ij}$} & (ours) & $\checkmark$ & $\checkmark$ &  \\
\textsf{injected power $P^d$} & \cite{schultzDetoursBasinStability2014, yangPowergridStabilityPredictions2021} & $\checkmark$ & $\checkmark$ &  \\
\textsf{$\max_j \cos(\Delta \phi)_{ij}$} & (ours) & $\checkmark$ & $\checkmark$ &  \\
\textsf{$\median_j \cos(\Delta \phi)_{ij}$} & (ours) & $\checkmark$ & $\checkmark$ &  \\
\textsf{$\min_j \cos(\Delta \phi)_{ij}$} & (ours) & $\checkmark$ & $\checkmark$ &  \\
\textsf{$\sum_j \cos(\Delta \phi)_{ij}$} & (ours) & $\checkmark$ & $\checkmark$ &  \\
\textsf{$\max_j (\Delta \phi)_{ij}$} & (ours) & $\checkmark$ & $\checkmark$ &  \\
\textsf{$\median_j (\Delta \phi)_{ij}$} & (ours) & $\checkmark$ & $\checkmark$ &  \\
\textsf{$\min_j (\Delta \phi)_{ij}$} & (ours) & $\checkmark$ & $\checkmark$ &  \\
\textsf{$\sum_j (\Delta \phi)_{ij}$} & (ours) & $\checkmark$ & $\checkmark$ &  \\
\textsf{$\max_j \text{Flow}_{ij}$} & (ours) & $\checkmark$ & $\checkmark$ &  \\
\textsf{$\median_j \text{Flow}_{ij}$} & (ours) & $\checkmark$ & $\checkmark$ &  \\
\textsf{$\min_j \text{Flow}_{ij}$} & (ours) & $\checkmark$ & $\checkmark$ &  \\
\textsf{node connected to maximally loaded line} & (ours) & $\checkmark$ & $\checkmark$ &  \\
\textsf{Damping $D$} & (ours) & $\checkmark$ &  &  \\
\textsf{is bulk} & \cite{nitzbonDecipheringImprintTopology2017} & $\square$ & $\checkmark$ & $\square$ \\
\textsf{is root node} & \cite{nitzbonDecipheringImprintTopology2017} & $\square$ & $\checkmark$ & $\square$ \\
\textsf{is sparse sprout node} & \cite{nitzbonDecipheringImprintTopology2017} & $\square$ & $\checkmark$ & $\square$ \\
\textsf{deg. assortativity} & (ours) & $\square$ & $\square$ & $\square$ \\
\textsf{is dense sprout} & \cite{nitzbonDecipheringImprintTopology2017} & $\square$ & $\square$ & $\square$ \\
\textsf{diameter} & \cite{feldLargedeviationsBasinStability2019} & $\square$ & $\square$ & $\square$ \\
\textsf{eigenratio} $\lambda_2/\lambda_N$ & \cite{zhouUniversalitySynchronizationWeighted2006, chenNetworkSynchronizabilityAnalysis2008} & $\square$ & $\square$ & $\square$ \\
\textsf{is inner tree node} & \cite{nitzbonDecipheringImprintTopology2017} & $\square$ & $\square$ & $\square$ \\
\textsf{inverse algebraic connectivity $1/\lambda_2$} & \cite{chenNetworkSynchronizabilityAnalysis2008, titzPredictingDynamicStability2024} & $\square$ & $\square$ & $\square$ \\
\textsf{Kirchhoff index} & \cite{tylooKeyPlayerProblem2019} & $\square$ & $\square$ & $\square$ \\
\textsf{proper leaf node} & \cite{nitzbonDecipheringImprintTopology2017} & $\square$ & $\square$ & $\square$ \\
\textsf{resistance distance Kirchhoff index} & \cite{tylooKeyPlayerProblem2019} & $\square$ & $\square$ & $\square$ \\
\textsf{transitivity} & \cite{feldLargedeviationsBasinStability2019} & $\square$ & $\square$ & $\square$ \\
\textsf{$P$ assortativity} & (ours) & $\square$ & $\square$ &  \\
\textsf{backup capacity} & (ours) & $\square$ & $\square$ &  \\
\textsf{grid connected if maximally loaded line fails} & (ours) & $\square$ & $\square$ &  \\
\textsf{maximal line load at operation point} & \cite{witthautCriticalLinksNonlocal2016, titzPredictingDynamicStability2024} & $\square$ & $\square$ &  \\
\textsf{maximal line load after failure (DC)} & (ours) & $\square$ & $\square$ &  \\
\textsf{power sign ratio} & \cite{dewenterLargedeviationPropertiesResilience2015,feldLargedeviationsBasinStability2019} & $\square$ & $\square$ &  \\
\textsf{size of grid} & (ours) & $\square$ &  &  \\
\textsf{universal Kuramoto order parameter} & \cite{schroderUniversalOrderParameter2017, feldLargedeviationsBasinStability2019} & $\square$ &  &  \\
\textsf{total inductance} & (ours) &  & $\checkmark$ &  \\
\textsf{reactive power $Q^d$} & (ours) &  & $\checkmark$ &  \\
\textsf{Dynamic type of node} & (ours) &  & $\checkmark$ &  \\
\textsf{$\sum_j \Im(Y_{ij})$} & (ours) &  & $\checkmark$ &  \\
\textsf{$\sum_j \Re(Y_{ij})$} & (ours) &  & $\checkmark$ &  \\
\textsf{Shunt inductance} & (ours) &  & $\checkmark$ &  \\
\end{longtable}

\section{Supplementary material: Summary of experimental results}
\label{sec:appendix_experiments_summary}

This section provides a concise overview of the experimental outcomes, highlighting selected results. Additional results are presented in the dedicated sections for each experiment. Specifically, \Cref{tab:WSKura summary} summarizes the WS-Kura results; \Cref{tab:VDP pearson selected nm,tab:VDP NetSciML} summarize the WS-VDP results; \Cref{tab:PGKura summary asymptotic,tab:PGKura summary transient} summarize the PG-Kura results; and \Cref{tab:NF summary} summarizes the PG-Real.

\begin{longtable}{p{3.5cm}|c|rrrr|rrrr}
\caption{WS-Kura: Stability predictions for the inertial Kuramoto-Sakaguchi model with heterogeneous parameters on Watts-Strogatz topologies. Pearson correlation/Theil's U (orange/purple, in \%) of selected network measures and $R^2$ score (blue, in \%) of NetSciML and GNNs trained on the ensemble with rewiring probability 0.5 and mean degree 6. HGBR (dyn. features) was trained using only the inhomogeneous nodal parameters as input ($P^d$, $D$ and total coupling strength at the node), HGBR (top. features) used only topological features.} \label{tab:WSKura summary} \\
\toprule
 &  & \multicolumn{4}{c}{Transient} & \multicolumn{4}{c}{Asymptotic} \\
 &  & 4 & 6 & 8 & 10 & 4 & 6 & 8 & 10 \\
Predictor &  &  &  &  &  &  &  &  &  \\
\midrule
\endfirsthead
\caption[]{\textbf{(continued)} WS-Kura: Stability predictions for the inertial Kuramoto-Sakaguchi model with heterogeneous parameters on Watts-Strogatz topologies. Pearson correlation/Theil's U (orange/purple, in \%) of selected network measures and $R^2$ score (blue, in \%) of NetSciML and GNNs trained on the ensemble with rewiring probability 0.5 and mean degree 6. HGBR (dyn. features) was trained using only the inhomogeneous nodal parameters as input ($P^d$, $D$ and total coupling strength at the node), HGBR (top. features) used only topological features.} \\
\toprule
 &  & \multicolumn{4}{c}{Transient} & \multicolumn{4}{c}{Asymptotic} \\
 &  & 4 & 6 & 8 & 10 & 4 & 6 & 8 & 10 \\
Predictor &  &  &  &  &  &  &  &  &  \\
\midrule
\endhead
\midrule
\multicolumn{10}{r}{Continued on next page} \\
\midrule
\endfoot
\bottomrule
\endlastfoot
\multirow[c]{5}{=}{\textsf{$\sum_j K_{ij}$}} & 0.1 & {\cellcolor[HTML]{FAEDDA}} \color[HTML]{000000} $-7$ & {\cellcolor[HTML]{F6F6F7}} \color[HTML]{000000} $1$ & {\cellcolor[HTML]{EDEDF3}} \color[HTML]{000000} $6$ & {\cellcolor[HTML]{EBECF3}} \color[HTML]{000000} $6$ & {\cellcolor[HTML]{F7F6F3}} \color[HTML]{000000} $-1$ & {\cellcolor[HTML]{FCE5C5}} \color[HTML]{000000} $-12$ & {\cellcolor[HTML]{FEDDAF}} \color[HTML]{000000} $-17$ & {\cellcolor[HTML]{FEE0B6}} \color[HTML]{000000} $-16$ \\*
 & 0.3 & {\cellcolor[HTML]{FBEAD2}} \color[HTML]{000000} $-9$ & {\cellcolor[HTML]{EFF0F4}} \color[HTML]{000000} $4$ & {\cellcolor[HTML]{E8E9F1}} \color[HTML]{000000} $8$ & {\cellcolor[HTML]{E7E8F1}} \color[HTML]{000000} $8$ & {\cellcolor[HTML]{F8F4EE}} \color[HTML]{000000} $-2$ & {\cellcolor[HTML]{FEDBAC}} \color[HTML]{000000} $-18$ & {\cellcolor[HTML]{FED7A2}} \color[HTML]{000000} $-20$ & {\cellcolor[HTML]{FEDAA9}} \color[HTML]{000000} $-19$ \\*
 & 0.5 & {\cellcolor[HTML]{FBEAD2}} \color[HTML]{000000} $-9$ & {\cellcolor[HTML]{EAEBF2}} \color[HTML]{000000} $7$ & {\cellcolor[HTML]{E5E7F0}} \color[HTML]{000000} $9$ & {\cellcolor[HTML]{E5E7F0}} \color[HTML]{000000} $9$ & {\cellcolor[HTML]{F9F0E4}} \color[HTML]{000000} $-5$ & {\cellcolor[HTML]{FED39C}} \color[HTML]{000000} $-21$ & {\cellcolor[HTML]{FED299}} \color[HTML]{000000} $-22$ & {\cellcolor[HTML]{FED59F}} \color[HTML]{000000} $-20$ \\*
 & 0.7 & {\cellcolor[HTML]{FAEEDC}} \color[HTML]{000000} $-6$ & {\cellcolor[HTML]{EAEBF2}} \color[HTML]{000000} $7$ & {\cellcolor[HTML]{E5E7F0}} \color[HTML]{000000} $9$ & {\cellcolor[HTML]{E7E8F1}} \color[HTML]{000000} $9$ & {\cellcolor[HTML]{FBEBD5}} \color[HTML]{000000} $-8$ & {\cellcolor[HTML]{FED299}} \color[HTML]{000000} $-22$ & {\cellcolor[HTML]{FED095}} \color[HTML]{000000} $-22$ & {\cellcolor[HTML]{FED59F}} \color[HTML]{000000} $-20$ \\
\end{longtable}

\begin{longtable}{p{3cm}rrrrrrrrr}
\caption{Pearson correlation (in \%) of selected network measures and stability index in the WS-VDP experiment.} \label{tab:VDP pearson selected nm} \\
\toprule
 &  & \multicolumn{4}{c}{Transient} & \multicolumn{4}{c}{Asymptotic} \\
 &  & 4 & 6 & 8 & 10 & 4 & 6 & 8 & 10 \\
Network Measure &  &  &  &  &  &  &  &  &  \\
\midrule
\endfirsthead
\caption[]{\textbf{(continued)} Pearson correlation (in \%) of selected network measures and stability index in the WS-VDP experiment.} \\
\toprule
 &  & \multicolumn{4}{c}{Transient} & \multicolumn{4}{c}{Asymptotic} \\
 &  & 4 & 6 & 8 & 10 & 4 & 6 & 8 & 10 \\
Network Measure &  &  &  &  &  &  &  &  &  \\
\midrule
\endhead
\midrule
\multicolumn{10}{r}{Continued on next page} \\
\midrule
\endfoot
\bottomrule
\endlastfoot
\multirow[c]{5}{=}{\textsf{node deg.}} & 0.1 & {\cellcolor[HTML]{D9DBEB}} \color[HTML]{000000} $15$ & {\cellcolor[HTML]{CCCBE3}} \color[HTML]{000000} $21$ & {\cellcolor[HTML]{EAEBF2}} \color[HTML]{000000} $6$ & {\cellcolor[HTML]{954807}} \color[HTML]{F1F1F1} $-73$ & {\cellcolor[HTML]{FDBD6E}} \color[HTML]{000000} $-30$ & {\cellcolor[HTML]{FDBC6B}} \color[HTML]{000000} $-30$ & {\cellcolor[HTML]{FCE6C8}} \color[HTML]{000000} $-11$ & {\cellcolor[HTML]{532687}} \color[HTML]{F1F1F1} $64$ \\*
 & 0.3 & {\cellcolor[HTML]{FED8A6}} \color[HTML]{000000} $-19$ & {\cellcolor[HTML]{E7E8F1}} \color[HTML]{000000} $9$ & {\cellcolor[HTML]{E68D23}} \color[HTML]{F1F1F1} $-44$ & {\cellcolor[HTML]{CE720F}} \color[HTML]{F1F1F1} $-54$ & {\cellcolor[HTML]{FDE5C3}} \color[HTML]{000000} $-13$ & {\cellcolor[HTML]{FDB965}} \color[HTML]{000000} $-31$ & {\cellcolor[HTML]{ABA3CD}} \color[HTML]{F1F1F1} $34$ & {\cellcolor[HTML]{F0F1F4}} \color[HTML]{000000} $4$ \\*
 & 0.5 & {\cellcolor[HTML]{FED299}} \color[HTML]{000000} $-21$ & {\cellcolor[HTML]{FBEBD5}} \color[HTML]{000000} $-8$ & {\cellcolor[HTML]{833D08}} \color[HTML]{F1F1F1} $-78$ & {\cellcolor[HTML]{A85206}} \color[HTML]{F1F1F1} $-67$ & {\cellcolor[HTML]{FED39C}} \color[HTML]{000000} $-21$ & {\cellcolor[HTML]{FDB965}} \color[HTML]{000000} $-32$ & {\cellcolor[HTML]{B7B1D5}} \color[HTML]{000000} $30$ & {\cellcolor[HTML]{F9F2E9}} \color[HTML]{000000} $-3$ \\*
 & 0.7 & {\cellcolor[HTML]{FED7A2}} \color[HTML]{000000} $-20$ & {\cellcolor[HTML]{FEDBAC}} \color[HTML]{000000} $-18$ & {\cellcolor[HTML]{7F3B08}} \color[HTML]{F1F1F1} $-82$ & {\cellcolor[HTML]{9E4C07}} \color[HTML]{F1F1F1} $-70$ & {\cellcolor[HTML]{FDBC6B}} \color[HTML]{000000} $-30$ & {\cellcolor[HTML]{FDBA68}} \color[HTML]{000000} $-31$ & {\cellcolor[HTML]{D1D1E6}} \color[HTML]{000000} $19$ & {\cellcolor[HTML]{F7F6F3}} \color[HTML]{000000} $-1$ \\
\end{longtable}

\begin{longtable}{crrrrrrrrrrrr}
\caption{$R^2$ (in \%) of prediction of stability by NetSciML and GNN in the WS-VDP experiment. NetSciML and GNN were trained on the degree $6$ and rewiring probability $0.5$ ensemble.} \label{tab:VDP NetSciML} \\
\toprule
 & \multicolumn{4}{c}{Transient} & \multicolumn{4}{c}{Asymptotic} & \multicolumn{4}{c}{Asymptotic (GNN)} \\
 & 4 & 6 & 8 & 10 & 4 & 6 & 8 & 10 & 4 & 6 & 8 & 10 \\
\midrule
\endfirsthead
\caption[]{\textbf{(continued)} $R^2$ (in \%) of prediction of stability by NetSciML and GNN in the WS-VDP experiment. NetSciML and GNN were trained on the degree $6$ and rewiring probability $0.5$ ensemble.} \\
\toprule
 & \multicolumn{4}{c}{Transient} & \multicolumn{4}{c}{Asymptotic} & \multicolumn{4}{c}{Asymptotic (GNN)} \\
 & 4 & 6 & 8 & 10 & 4 & 6 & 8 & 10 & 4 & 6 & 8 & 10 \\
\midrule
\endhead
\midrule
\multicolumn{13}{r}{Continued on next page} \\
\midrule
\endfoot
\bottomrule
\endlastfoot
0.1 & {\cellcolor[HTML]{F7FBFF}} \color[HTML]{000000} \color{gray} $-44$ & {\cellcolor[HTML]{F7FBFF}} \color[HTML]{000000} \color{gray} $\ll 0$ & {\cellcolor[HTML]{F7FBFF}} \color[HTML]{000000} \color{gray} $\ll 0$ & {\cellcolor[HTML]{EFF6FC}} \color[HTML]{000000} $4$ & {\cellcolor[HTML]{F7FBFF}} \color[HTML]{000000} \color{gray} $\ll 0$ & {\cellcolor[HTML]{F7FBFF}} \color[HTML]{000000} \color{gray} $-40$ & {\cellcolor[HTML]{F7FBFF}} \color[HTML]{000000} \color{gray} $\ll 0$ & {\cellcolor[HTML]{F7FBFF}} \color[HTML]{000000} \color{gray} $\ll 0$ & {\cellcolor[HTML]{F7FBFF}} \color[HTML]{000000} \color{gray} $\ll 0$ & {\cellcolor[HTML]{F7FBFF}} \color[HTML]{000000} \color{gray} $-92$ & {\cellcolor[HTML]{F7FBFF}} \color[HTML]{000000} \color{gray} $-8$ & {\cellcolor[HTML]{F7FBFF}} \color[HTML]{000000} \color{gray} $\ll 0$ \\
0.3 & {\cellcolor[HTML]{F7FBFF}} \color[HTML]{000000} \color{gray} $-7$ & {\cellcolor[HTML]{CFE1F2}} \color[HTML]{000000} $20$ & {\cellcolor[HTML]{C3DAEE}} \color[HTML]{000000} $26$ & {\cellcolor[HTML]{F7FBFF}} \color[HTML]{000000} \color{gray} $\ll 0$ & {\cellcolor[HTML]{F7FBFF}} \color[HTML]{000000} \color{gray} $-17$ & {\cellcolor[HTML]{BCD7EB}} \color[HTML]{000000} $28$ & {\cellcolor[HTML]{F7FBFF}} \color[HTML]{000000} \color{gray} $\ll 0$ & {\cellcolor[HTML]{F7FBFF}} \color[HTML]{000000} \color{gray} $\ll 0$ & {\cellcolor[HTML]{F7FBFF}} \color[HTML]{000000} \color{gray} $\ll 0$ & {\cellcolor[HTML]{E7F1FA}} \color[HTML]{000000} $8$ & {\cellcolor[HTML]{F7FBFF}} \color[HTML]{000000} \color{gray} $\ll 0$ & {\cellcolor[HTML]{F7FBFF}} \color[HTML]{000000} \color{gray} $\ll 0$ \\
\end{longtable}

\begin{longtable}{p{3.5cm}|rrrrrrr}
\caption{PG-Kura: Prediction of asymptotic stability for the inertial Kuramoto model on power grid like topologies. Pearson correlation (orange/purple, in \%) and $R^2$ score (blue, in \%) of selected network measures and of NetSciML and GNN models trained on the 100 node ensemble. NetSciML and GNN are trained on 7000 networks, NetSciML (fewer grids) on 70 networks, NetSciML (no PF features) without network measures derived from the synchronous state.} \label{tab:PGKura summary asymptotic} \\
\toprule
 & 20 & 100 & DE & FR & GB & ES & TX \\
Predictor &  &  &  &  &  &  &  \\
\midrule
\endfirsthead
\caption[]{\textbf{(continued)} PG-Kura: Prediction of asymptotic stability for the inertial Kuramoto model on power grid like topologies. Pearson correlation (orange/purple, in \%) and $R^2$ score (blue, in \%) of selected network measures and of NetSciML and GNN models trained on the 100 node ensemble. NetSciML and GNN are trained on 7000 networks, NetSciML (fewer grids) on 70 networks, NetSciML (no PF features) without network measures derived from the synchronous state.} \\
\toprule
 & 20 & 100 & DE & FR & GB & ES & TX \\
Predictor &  &  &  &  &  &  &  \\
\midrule
\endhead
\midrule
\multicolumn{8}{r}{Continued on next page} \\
\midrule
\endfoot
\bottomrule
\endlastfoot
\textsf{node deg.} & {\cellcolor[HTML]{664697}} \color[HTML]{F1F1F1} $57$ & {\cellcolor[HTML]{7D6DA9}} \color[HTML]{F1F1F1} $49$ & {\cellcolor[HTML]{AFA8D0}} \color[HTML]{000000} $33$ & {\cellcolor[HTML]{9287B9}} \color[HTML]{F1F1F1} $42$ & {\cellcolor[HTML]{725BA1}} \color[HTML]{F1F1F1} $53$ & {\cellcolor[HTML]{644395}} \color[HTML]{F1F1F1} $58$ & {\cellcolor[HTML]{664697}} \color[HTML]{F1F1F1} $57$ \\
\textsf{average neighbor deg.} & {\cellcolor[HTML]{CCCBE3}} \color[HTML]{000000} $21$ & {\cellcolor[HTML]{B3ACD2}} \color[HTML]{000000} $32$ & {\cellcolor[HTML]{F8F5F1}} \color[HTML]{000000} $-2$ & {\cellcolor[HTML]{C6C4DF}} \color[HTML]{000000} $24$ & {\cellcolor[HTML]{CCCBE3}} \color[HTML]{000000} $21$ & {\cellcolor[HTML]{8A7EB3}} \color[HTML]{F1F1F1} $45$ & {\cellcolor[HTML]{D57811}} \color[HTML]{F1F1F1} $-52$ \\
\textsf{eigenvector cent.} & {\cellcolor[HTML]{644395}} \color[HTML]{F1F1F1} $58$ & {\cellcolor[HTML]{8C80B5}} \color[HTML]{F1F1F1} $44$ & {\cellcolor[HTML]{DDDFED}} \color[HTML]{000000} $13$ & {\cellcolor[HTML]{A198C5}} \color[HTML]{F1F1F1} $37$ & {\cellcolor[HTML]{A198C5}} \color[HTML]{F1F1F1} $37$ & {\cellcolor[HTML]{CECDE4}} \color[HTML]{000000} $20$ & {\cellcolor[HTML]{F8F4EE}} \color[HTML]{000000} $-2$ \\
\textsf{$A^3 P^d$} & {\cellcolor[HTML]{F6F6F7}} \color[HTML]{000000} $0$ & {\cellcolor[HTML]{F6F6F7}} \color[HTML]{000000} $0$ & {\cellcolor[HTML]{F8F4EE}} \color[HTML]{000000} $-2$ & {\cellcolor[HTML]{FDC175}} \color[HTML]{000000} $-28$ & {\cellcolor[HTML]{FDE5C3}} \color[HTML]{000000} $-13$ & {\cellcolor[HTML]{F5F5F6}} \color[HTML]{000000} $1$ & {\cellcolor[HTML]{F4F4F6}} \color[HTML]{000000} $2$ \\
\end{longtable}

\begin{longtable}{p{3.5cm}|rrrrrrr}
\caption{PG-Kura: Prediction of transient stability for the inertial Kuramoto model on power grid like topologies. Pearson correlation (orange/purple, in \%) and $R^2$ score (blue, in \%) of selected network measures and of NetSciML and GNN models trained on the 100 node ensemble. NetSciML and GNN are trained on 7000 networks, NetSciML (fewer grids) on 70 networks, NetSciML (no PF features) without network measures derived from the synchronous state.} \label{tab:PGKura summary transient} \\
\toprule
 & 20 & 100 & DE & FR & GB & ES & TX \\
Predictor &  &  &  &  &  &  &  \\
\midrule
\endfirsthead
\caption[]{\textbf{(continued)} PG-Kura: Prediction of transient stability for the inertial Kuramoto model on power grid like topologies. Pearson correlation (orange/purple, in \%) and $R^2$ score (blue, in \%) of selected network measures and of NetSciML and GNN models trained on the 100 node ensemble. NetSciML and GNN are trained on 7000 networks, NetSciML (fewer grids) on 70 networks, NetSciML (no PF features) without network measures derived from the synchronous state.} \\
\toprule
 & 20 & 100 & DE & FR & GB & ES & TX \\
Predictor &  &  &  &  &  &  &  \\
\midrule
\endhead
\midrule
\multicolumn{8}{r}{Continued on next page} \\
\midrule
\endfoot
\bottomrule
\endlastfoot
\textsf{node deg.} & {\cellcolor[HTML]{2D004B}} \color[HTML]{F1F1F1} $91$ & {\cellcolor[HTML]{2D004B}} \color[HTML]{F1F1F1} $94$ & {\cellcolor[HTML]{2D004B}} \color[HTML]{F1F1F1} $95$ & {\cellcolor[HTML]{2D004B}} \color[HTML]{F1F1F1} $95$ & {\cellcolor[HTML]{2D004B}} \color[HTML]{F1F1F1} $99$ & {\cellcolor[HTML]{2D004B}} \color[HTML]{F1F1F1} $98$ & {\cellcolor[HTML]{2D004B}} \color[HTML]{F1F1F1} $93$ \\
\textsf{average neighbor deg.} & {\cellcolor[HTML]{FCE8CD}} \color[HTML]{000000} $-10$ & {\cellcolor[HTML]{EDEDF3}} \color[HTML]{000000} $5$ & {\cellcolor[HTML]{FED095}} \color[HTML]{000000} $-22$ & {\cellcolor[HTML]{FDC47B}} \color[HTML]{000000} $-27$ & {\cellcolor[HTML]{FAEDDA}} \color[HTML]{000000} $-7$ & {\cellcolor[HTML]{C5C2DE}} \color[HTML]{000000} $24$ & {\cellcolor[HTML]{FBEBD5}} \color[HTML]{000000} $-9$ \\
\textsf{eigenvector cent.} & {\cellcolor[HTML]{380B5C}} \color[HTML]{F1F1F1} $75$ & {\cellcolor[HTML]{4C1F7B}} \color[HTML]{F1F1F1} $67$ & {\cellcolor[HTML]{8C80B5}} \color[HTML]{F1F1F1} $44$ & {\cellcolor[HTML]{8477AF}} \color[HTML]{F1F1F1} $46$ & {\cellcolor[HTML]{70589F}} \color[HTML]{F1F1F1} $53$ & {\cellcolor[HTML]{9990BF}} \color[HTML]{F1F1F1} $40$ & {\cellcolor[HTML]{9B92C1}} \color[HTML]{F1F1F1} $39$ \\
\textsf{$A^3 P^d$} & {\cellcolor[HTML]{F7F6F3}} \color[HTML]{000000} $-1$ & {\cellcolor[HTML]{F7F7F6}} \color[HTML]{000000} $0$ & {\cellcolor[HTML]{F6F6F7}} \color[HTML]{000000} $0$ & {\cellcolor[HTML]{EF9E3C}} \color[HTML]{000000} $-39$ & {\cellcolor[HTML]{FCE8CD}} \color[HTML]{000000} $-10$ & {\cellcolor[HTML]{F4F4F6}} \color[HTML]{000000} $1$ & {\cellcolor[HTML]{EBECF3}} \color[HTML]{000000} $6$ \\
\end{longtable}

\begin{longtable}{p{3.5cm}|rrrrr|rrrrr}
\caption{PG-Real: Stability predictions for realistic power grid dynamics on synthetic topologies. Pearson correlation (orange/purple, in \%) of selected network measures and and $R^2$ score (blue, in \%) of NetSciML and GNN models trained on the ensemble with 70-80 node grids.} \label{tab:NF summary} \\
\toprule
 & \multicolumn{5}{c}{Asymptotic} & \multicolumn{5}{c}{Transient} \\
 & 70-80 & 128 & 256 & 512 & 1024 & 70-80 & 128 & 256 & 512 & 1024 \\
Predictor &  &  &  &  &  &  &  &  &  &  \\
\midrule
\endfirsthead
\caption[]{\textbf{(continued)} PG-Real: Stability predictions for realistic power grid dynamics on synthetic topologies. Pearson correlation (orange/purple, in \%) of selected network measures and and $R^2$ score (blue, in \%) of NetSciML and GNN models trained on the ensemble with 70-80 node grids.} \\
\toprule
 & \multicolumn{5}{c}{Asymptotic} & \multicolumn{5}{c}{Transient} \\
 & 70-80 & 128 & 256 & 512 & 1024 & 70-80 & 128 & 256 & 512 & 1024 \\
Predictor &  &  &  &  &  &  &  &  &  &  \\
\midrule
\endhead
\midrule
\multicolumn{11}{r}{Continued on next page} \\
\midrule
\endfoot
\bottomrule
\endlastfoot
\textsf{$\max_j K_{ij}$} & {\cellcolor[HTML]{F5F5F6}} \color[HTML]{000000} $1$ & {\cellcolor[HTML]{F8F4EE}} \color[HTML]{000000} $-2$ & {\cellcolor[HTML]{A39BC7}} \color[HTML]{F1F1F1} $36$ & {\cellcolor[HTML]{F6F6F7}} \color[HTML]{000000} $0$ & {\cellcolor[HTML]{F2F2F5}} \color[HTML]{000000} $3$ & {\cellcolor[HTML]{F5F5F6}} \color[HTML]{000000} $1$ & {\cellcolor[HTML]{FBE9CF}} \color[HTML]{000000} $-10$ & {\cellcolor[HTML]{EFF0F4}} \color[HTML]{000000} $4$ & {\cellcolor[HTML]{F4F4F6}} \color[HTML]{000000} $2$ & {\cellcolor[HTML]{E7E8F1}} \color[HTML]{000000} $8$ \\
\textsf{$\max_j \cos(\Delta \phi)_{ij}$} & {\cellcolor[HTML]{FAEEDF}} \color[HTML]{000000} $-6$ & {\cellcolor[HTML]{FBEBD5}} \color[HTML]{000000} $-8$ & {\cellcolor[HTML]{FAECD7}} \color[HTML]{000000} $-8$ & {\cellcolor[HTML]{FEDBAC}} \color[HTML]{000000} $-18$ & {\cellcolor[HTML]{FCE7CA}} \color[HTML]{000000} $-11$ & {\cellcolor[HTML]{EBECF3}} \color[HTML]{000000} $6$ & {\cellcolor[HTML]{E4E5F0}} \color[HTML]{000000} $10$ & {\cellcolor[HTML]{EBECF3}} \color[HTML]{000000} $6$ & {\cellcolor[HTML]{FECD8F}} \color[HTML]{000000} $-24$ & {\cellcolor[HTML]{F7AC52}} \color[HTML]{000000} $-35$ \\
\textsf{Dynamic type of node} & {\cellcolor[HTML]{CFCFE5}} \color[HTML]{000000} $20$ & {\cellcolor[HTML]{C0BDDB}} \color[HTML]{000000} $26$ & {\cellcolor[HTML]{CECDE4}} \color[HTML]{000000} $20$ & {\cellcolor[HTML]{C0BDDB}} \color[HTML]{000000} $26$ & {\cellcolor[HTML]{C0BDDB}} \color[HTML]{000000} $26$ & {\cellcolor[HTML]{320552}} \color[HTML]{F1F1F1} $78$ & {\cellcolor[HTML]{360959}} \color[HTML]{F1F1F1} $76$ & {\cellcolor[HTML]{350857}} \color[HTML]{F1F1F1} $76$ & {\cellcolor[HTML]{350857}} \color[HTML]{F1F1F1} $77$ & {\cellcolor[HTML]{360959}} \color[HTML]{F1F1F1} $76$ \\
\end{longtable}

\FloatBarrier

\section{Supplementary material on WS-Kura experiments}
\label{sec:appendix_experiments_WSKura}
We analyze Watts–Strogatz ensembles with mean degree $d \in \{4,6,8,10\}$ and rewiring probability $p_r \in \{0.1,0.3,0.5,0.7,0.9\}$, governed by the inertial Kuramoto–Sakaguchi model with strongly heterogeneous parameters. The varying parameters are the coupling strengths $K_{ij}$, phase shift $\alpha$, damping $D$, and injected power $P^d$. Asymptotic stability is quantified via single-node basin stability, defined as the fraction of perturbations for which the system returns to synchrony with frequencies within narrow bounds, while transient stability is assessed using a synchronization norm that measures the magnitude of desynchronization during recovery.

\Cref{tab:WSKura all correlations} reports Pearson correlations for the network measures in WS-Kura, while \Cref{tab:WSKura all TU} reports Theil’s $U$. For a selected subset of network measures, \Cref{tab:WSKura pearson selected nm} provides the corresponding Pearson correlations.


\begin{longtable}{p{3cm}cccccccccc}
\caption{Pearson correlation (in \%) of network measures and stability index in the WS-Kura experiment.} \label{tab:WSKura all correlations} \\
\toprule
 & \multicolumn{5}{c}{Asymptotic} & \multicolumn{5}{c}{Transient} \\
 & 6 0.5 & 4 0.3 & 8 0.3 & 4 0.7 & 8 0.7 & 6 0.5 & 4 0.3 & 8 0.3 & 4 0.7 & 8 0.7 \\
Network Measure &  &  &  &  &  &  &  &  &  &  \\
\midrule
\endfirsthead
\caption[]{\textbf{(continued)} Pearson correlation (in \%) of network measures and stability index in the WS-Kura experiment.} \\
\toprule
 & \multicolumn{5}{c}{Asymptotic} & \multicolumn{5}{c}{Transient} \\
 & 6 0.5 & 4 0.3 & 8 0.3 & 4 0.7 & 8 0.7 & 6 0.5 & 4 0.3 & 8 0.3 & 4 0.7 & 8 0.7 \\
Network Measure &  &  &  &  &  &  &  &  &  &  \\
\midrule
\endhead
\midrule
\multicolumn{11}{r}{Continued on next page} \\
\midrule
\endfoot
\bottomrule
\endlastfoot
\textsf{$\sum_j \cos(\Delta \phi)_{ij}$} & {\cellcolor[HTML]{FECF92}} \color[HTML]{000000} $-23$ & {\cellcolor[HTML]{FBE9CF}} \color[HTML]{000000} $-9$ & {\cellcolor[HTML]{FED39C}} \color[HTML]{000000} $-21$ & {\cellcolor[HTML]{FDE4C0}} \color[HTML]{000000} $-13$ & {\cellcolor[HTML]{FECD8F}} \color[HTML]{000000} $-23$ & {\cellcolor[HTML]{E5E7F0}} \color[HTML]{000000} $9$ & {\cellcolor[HTML]{F8F3EC}} \color[HTML]{000000} $-3$ & {\cellcolor[HTML]{E5E7F0}} \color[HTML]{000000} $9$ & {\cellcolor[HTML]{F8F4EE}} \color[HTML]{000000} $-2$ & {\cellcolor[HTML]{E4E5F0}} \color[HTML]{000000} $10$ \\
\cline{1-11}
\textsf{$\min_j \cos(\Delta \phi)_{ij}$} & {\cellcolor[HTML]{FED39C}} \color[HTML]{000000} $-21$ & {\cellcolor[HTML]{FCE7CA}} \color[HTML]{000000} $-11$ & {\cellcolor[HTML]{FEDEB3}} \color[HTML]{000000} $-16$ & {\cellcolor[HTML]{FEDAA9}} \color[HTML]{000000} $-19$ & {\cellcolor[HTML]{FED8A6}} \color[HTML]{000000} $-19$ & {\cellcolor[HTML]{E8E9F1}} \color[HTML]{000000} $8$ & {\cellcolor[HTML]{F8F5F1}} \color[HTML]{000000} $-2$ & {\cellcolor[HTML]{E9EAF2}} \color[HTML]{000000} $7$ & {\cellcolor[HTML]{F5F5F6}} \color[HTML]{000000} $1$ & {\cellcolor[HTML]{E5E7F0}} \color[HTML]{000000} $9$ \\
\cline{1-11}
\textsf{$\sum_j K_{ij}$} & {\cellcolor[HTML]{FED39C}} \color[HTML]{000000} $-21$ & {\cellcolor[HTML]{F8F4EE}} \color[HTML]{000000} $-2$ & {\cellcolor[HTML]{FED7A2}} \color[HTML]{000000} $-20$ & {\cellcolor[HTML]{FBEBD5}} \color[HTML]{000000} $-8$ & {\cellcolor[HTML]{FED095}} \color[HTML]{000000} $-22$ & {\cellcolor[HTML]{EAEBF2}} \color[HTML]{000000} $7$ & {\cellcolor[HTML]{FBEAD2}} \color[HTML]{000000} $-9$ & {\cellcolor[HTML]{E8E9F1}} \color[HTML]{000000} $8$ & {\cellcolor[HTML]{FAEEDC}} \color[HTML]{000000} $-6$ & {\cellcolor[HTML]{E5E7F0}} \color[HTML]{000000} $9$ \\
\end{longtable}

\begin{longtable}{p{3cm}cccccccccc}
\caption{Theil's U (in \%) of network measures and stability index in the WS-Kura experiment.} \label{tab:WSKura all TU} \\
\toprule
 & \multicolumn{5}{c}{Asymptotic} & \multicolumn{5}{c}{Transient} \\
 & 6 0.5 & 4 0.3 & 8 0.3 & 4 0.7 & 8 0.7 & 6 0.5 & 4 0.3 & 8 0.3 & 4 0.7 & 8 0.7 \\
Network Measure &  &  &  &  &  &  &  &  &  &  \\
\midrule
\endfirsthead
\caption[]{\textbf{(continued)} Theil's U (in \%) of network measures and stability index in the WS-Kura experiment.} \\
\toprule
 & \multicolumn{5}{c}{Asymptotic} & \multicolumn{5}{c}{Transient} \\
 & 6 0.5 & 4 0.3 & 8 0.3 & 4 0.7 & 8 0.7 & 6 0.5 & 4 0.3 & 8 0.3 & 4 0.7 & 8 0.7 \\
Network Measure &  &  &  &  &  &  &  &  &  &  \\
\midrule
\endhead
\midrule
\multicolumn{11}{r}{Continued on next page} \\
\midrule
\endfoot
\bottomrule
\endlastfoot
\textsf{is bulk} & {\cellcolor[HTML]{F6F6F7}} \color[HTML]{000000} $0$ & {\cellcolor[HTML]{F6F6F7}} \color[HTML]{000000} $0$ & {\cellcolor[HTML]{F6F6F7}} \color[HTML]{000000} $0$ & {\cellcolor[HTML]{F6F6F7}} \color[HTML]{000000} $0$ & {\cellcolor[HTML]{F6F6F7}} \color[HTML]{000000} $0$ & {\cellcolor[HTML]{F6F6F7}} \color[HTML]{000000} $0$ & {\cellcolor[HTML]{F6F6F7}} \color[HTML]{000000} $0$ & {\cellcolor[HTML]{F6F6F7}} \color[HTML]{000000} $0$ & {\cellcolor[HTML]{F6F6F7}} \color[HTML]{000000} $0$ & {\cellcolor[HTML]{F6F6F7}} \color[HTML]{000000} $0$ \\
\cline{1-11}
\textsf{grid connected if maximally loaded line fails} & {\cellcolor[HTML]{F6F6F7}} \color[HTML]{000000} $0$ & {\cellcolor[HTML]{F6F6F7}} \color[HTML]{000000} $0$ & {\cellcolor[HTML]{F6F6F7}} \color[HTML]{000000} $0$ & {\cellcolor[HTML]{F6F6F7}} \color[HTML]{000000} $0$ & {\cellcolor[HTML]{F6F6F7}} \color[HTML]{000000} $0$ & {\cellcolor[HTML]{F6F6F7}} \color[HTML]{000000} $0$ & {\cellcolor[HTML]{F6F6F7}} \color[HTML]{000000} $0$ & {\cellcolor[HTML]{F6F6F7}} \color[HTML]{000000} $0$ & {\cellcolor[HTML]{F6F6F7}} \color[HTML]{000000} $0$ & {\cellcolor[HTML]{F6F6F7}} \color[HTML]{000000} $0$ \\
\cline{1-11}
\textsf{is dense sprout} & {\cellcolor[HTML]{F6F6F7}} \color[HTML]{000000} $0$ & {\cellcolor[HTML]{F6F6F7}} \color[HTML]{000000} $0$ & {\cellcolor[HTML]{F6F6F7}} \color[HTML]{000000} $0$ & {\cellcolor[HTML]{F6F6F7}} \color[HTML]{000000} $0$ & {\cellcolor[HTML]{F6F6F7}} \color[HTML]{000000} $0$ & {\cellcolor[HTML]{F6F6F7}} \color[HTML]{000000} $0$ & {\cellcolor[HTML]{F6F6F7}} \color[HTML]{000000} $0$ & {\cellcolor[HTML]{F6F6F7}} \color[HTML]{000000} $0$ & {\cellcolor[HTML]{F6F6F7}} \color[HTML]{000000} $0$ & {\cellcolor[HTML]{F6F6F7}} \color[HTML]{000000} $0$ \\
\end{longtable}

\begin{longtable}{p{3cm}|c|rrrr|rrrr}
\caption{Pearson correlation and Theil's U (in \%) between network measure and stability for selected measures in the WS-Kura experiment.} \label{tab:WSKura pearson selected nm} \\
\toprule
 &  & \multicolumn{4}{c}{Transient} & \multicolumn{4}{c}{Asymptotic} \\
 &  & 4 & 6 & 8 & 10 & 4 & 6 & 8 & 10 \\
Network Measure &  &  &  &  &  &  &  &  &  \\
\midrule
\endfirsthead
\caption[]{\textbf{(continued)} Pearson correlation and Theil's U (in \%) between network measure and stability for selected measures in the WS-Kura experiment.} \\
\toprule
 &  & \multicolumn{4}{c}{Transient} & \multicolumn{4}{c}{Asymptotic} \\
 &  & 4 & 6 & 8 & 10 & 4 & 6 & 8 & 10 \\
Network Measure &  &  &  &  &  &  &  &  &  \\
\midrule
\endhead
\midrule
\multicolumn{10}{r}{Continued on next page} \\
\midrule
\endfoot
\bottomrule
\endlastfoot
\multirow[c]{5}{=}{\textsf{$\sum_j K_{ij}$}} & 0.1 & {\cellcolor[HTML]{FAEDDA}} \color[HTML]{000000} $-7$ & {\cellcolor[HTML]{F6F6F7}} \color[HTML]{000000} $1$ & {\cellcolor[HTML]{EDEDF3}} \color[HTML]{000000} $6$ & {\cellcolor[HTML]{EBECF3}} \color[HTML]{000000} $6$ & {\cellcolor[HTML]{F7F6F3}} \color[HTML]{000000} $-1$ & {\cellcolor[HTML]{FCE5C5}} \color[HTML]{000000} $-12$ & {\cellcolor[HTML]{FEDDAF}} \color[HTML]{000000} $-17$ & {\cellcolor[HTML]{FEE0B6}} \color[HTML]{000000} $-16$ \\*
 & 0.3 & {\cellcolor[HTML]{FBEAD2}} \color[HTML]{000000} $-9$ & {\cellcolor[HTML]{EFF0F4}} \color[HTML]{000000} $4$ & {\cellcolor[HTML]{E8E9F1}} \color[HTML]{000000} $8$ & {\cellcolor[HTML]{E7E8F1}} \color[HTML]{000000} $8$ & {\cellcolor[HTML]{F8F4EE}} \color[HTML]{000000} $-2$ & {\cellcolor[HTML]{FEDBAC}} \color[HTML]{000000} $-18$ & {\cellcolor[HTML]{FED7A2}} \color[HTML]{000000} $-20$ & {\cellcolor[HTML]{FEDAA9}} \color[HTML]{000000} $-19$ \\*
 & 0.5 & {\cellcolor[HTML]{FBEAD2}} \color[HTML]{000000} $-9$ & {\cellcolor[HTML]{EAEBF2}} \color[HTML]{000000} $7$ & {\cellcolor[HTML]{E5E7F0}} \color[HTML]{000000} $9$ & {\cellcolor[HTML]{E5E7F0}} \color[HTML]{000000} $9$ & {\cellcolor[HTML]{F9F0E4}} \color[HTML]{000000} $-5$ & {\cellcolor[HTML]{FED39C}} \color[HTML]{000000} $-21$ & {\cellcolor[HTML]{FED299}} \color[HTML]{000000} $-22$ & {\cellcolor[HTML]{FED59F}} \color[HTML]{000000} $-20$ \\*
 & 0.7 & {\cellcolor[HTML]{FAEEDC}} \color[HTML]{000000} $-6$ & {\cellcolor[HTML]{EAEBF2}} \color[HTML]{000000} $7$ & {\cellcolor[HTML]{E5E7F0}} \color[HTML]{000000} $9$ & {\cellcolor[HTML]{E7E8F1}} \color[HTML]{000000} $9$ & {\cellcolor[HTML]{FBEBD5}} \color[HTML]{000000} $-8$ & {\cellcolor[HTML]{FED299}} \color[HTML]{000000} $-22$ & {\cellcolor[HTML]{FED095}} \color[HTML]{000000} $-22$ & {\cellcolor[HTML]{FED59F}} \color[HTML]{000000} $-20$ \\
\end{longtable}

\begin{longtable}{p{3cm}rrrrrrrrr}
\caption{$R^2$ (in \%) of prediction of stability by NetSciML in the WS-Kura experiment. NetSciML was trained on the degree $6$ and rewiring probability $0.5$ ensemble. HGBR (dyn. features) was trained using only the inhomogeneous nodal parameters as input ($P^d$, $D$ and total coupling strength at the node), HGBR (top. features) used only topological features.} \label{tab:WSKura NetSciML} \\
\toprule
 &  & \multicolumn{4}{c}{Transient} & \multicolumn{4}{c}{Asymptotic} \\
 &  & 4 & 6 & 8 & 10 & 4 & 6 & 8 & 10 \\
\midrule
\endfirsthead
\caption[]{\textbf{(continued)} $R^2$ (in \%) of prediction of stability by NetSciML in the WS-Kura experiment. NetSciML was trained on the degree $6$ and rewiring probability $0.5$ ensemble. HGBR (dyn. features) was trained using only the inhomogeneous nodal parameters as input ($P^d$, $D$ and total coupling strength at the node), HGBR (top. features) used only topological features.} \\
\toprule
 &  & \multicolumn{4}{c}{Transient} & \multicolumn{4}{c}{Asymptotic} \\
 &  & 4 & 6 & 8 & 10 & 4 & 6 & 8 & 10 \\
\midrule
\endhead
\midrule
\multicolumn{10}{r}{Continued on next page} \\
\midrule
\endfoot
\bottomrule
\endlastfoot
\multirow[c]{5}{=}{NetSciML} & 0.1 & {\cellcolor[HTML]{F7FBFF}} \color[HTML]{000000} \color{gray} $-17$ & {\cellcolor[HTML]{F7FBFF}} \color[HTML]{000000} \color{gray} $-4$ & {\cellcolor[HTML]{519CCC}} \color[HTML]{F1F1F1} $58$ & {\cellcolor[HTML]{08478D}} \color[HTML]{F1F1F1} $91$ & {\cellcolor[HTML]{F7FBFF}} \color[HTML]{000000} \color{gray} $-40$ & {\cellcolor[HTML]{9CC9E1}} \color[HTML]{000000} $38$ & {\cellcolor[HTML]{1562A9}} \color[HTML]{F1F1F1} $81$ & {\cellcolor[HTML]{125DA6}} \color[HTML]{F1F1F1} $83$ \\*
 & 0.3 & {\cellcolor[HTML]{F7FBFF}} \color[HTML]{000000} \color{gray} $-18$ & {\cellcolor[HTML]{CDE0F1}} \color[HTML]{000000} $21$ & {\cellcolor[HTML]{08468B}} \color[HTML]{F1F1F1} $92$ & {\cellcolor[HTML]{084C95}} \color[HTML]{F1F1F1} $89$ & {\cellcolor[HTML]{E1EDF8}} \color[HTML]{000000} $11$ & {\cellcolor[HTML]{2C7CBA}} \color[HTML]{F1F1F1} $71$ & {\cellcolor[HTML]{084D96}} \color[HTML]{F1F1F1} $89$ & {\cellcolor[HTML]{1764AB}} \color[HTML]{F1F1F1} $80$ \\*
 & 0.5 & {\cellcolor[HTML]{F7FBFF}} \color[HTML]{000000} \color{gray} $\ll 0$ & {\cellcolor[HTML]{2272B6}} \color[HTML]{F1F1F1} $75$ & {\cellcolor[HTML]{08478D}} \color[HTML]{F1F1F1} $91$ & {\cellcolor[HTML]{084C95}} \color[HTML]{F1F1F1} $89$ & {\cellcolor[HTML]{D6E5F4}} \color[HTML]{000000} $17$ & {\cellcolor[HTML]{1C6BB0}} \color[HTML]{F1F1F1} $77$ & {\cellcolor[HTML]{084F99}} \color[HTML]{F1F1F1} $88$ & {\cellcolor[HTML]{1C6BB0}} \color[HTML]{F1F1F1} $77$ \\*
 & 0.7 & {\cellcolor[HTML]{F7FBFF}} \color[HTML]{000000} \color{gray} $\ll 0$ & {\cellcolor[HTML]{7FB9DA}} \color[HTML]{000000} $45$ & {\cellcolor[HTML]{08478D}} \color[HTML]{F1F1F1} $91$ & {\cellcolor[HTML]{084990}} \color[HTML]{F1F1F1} $90$ & {\cellcolor[HTML]{DCEAF6}} \color[HTML]{000000} $14$ & {\cellcolor[HTML]{1E6DB2}} \color[HTML]{F1F1F1} $76$ & {\cellcolor[HTML]{08509B}} \color[HTML]{F1F1F1} $88$ & {\cellcolor[HTML]{1F6EB3}} \color[HTML]{F1F1F1} $76$ \\
\end{longtable}

In \Cref{tab:WSKura summary}, correlations vary systematically across two regimes: a sparse/regular regime and a denser more random regime. A nonlinear combination of network measures (NetSciML), trained at $p_r=0.5$ and $d=6$, generalizes well within the dense regime but degrades markedly in the sparse regime. NetSciML attains strong performance using only dynamical features: Theil’s $U$ for the injected power $P^d$ indicates high informativeness in the dense regime but substantially reduced informativeness in the sparse regime. Using only three dynamical inputs, $P^d$, $D$, and $\sum_j K_{ij}$, NetSciML already matches—or sometimes exceeds—the performance obtained with the full set of network descriptors. By contrast, in the sparse regime the universal Kuramoto order parameter correlates strongly with stability, indicating that properties of the synchronous state become increasingly predictive there.

\begin{longtable}{c|rrrr|rrrr|rrrr|rrrr}
\caption{$R^2$ (in \%) of prediction of asymptotic stability by GNN and NetSciML in the WS-Kura experiment. GNN and NetSciML were trained on the degree $6$ and rewiring probability $0.5$ ensemble. HGBR (dyn. features) was trained using only the inhomogeneous nodal parameters as input ($P^d$, $D$ and total coupling strength at the node), HGBR (top. features) used only topological features.} \label{tab:WSKura NetSciML GNN} \\
\toprule
 & \multicolumn{4}{c}{NetSciML} & \multicolumn{4}{c}{\makecell[l]{NetSciML \\(dyn. features)}} & \multicolumn{4}{c}{\makecell[l]{NetSciML\\(top. features)}} & \multicolumn{4}{c}{GNN} \\
 & 4 & 6 & 8 & 10 & 4 & 6 & 8 & 10 & 4 & 6 & 8 & 10 & 4 & 6 & 8 & 10 \\
\midrule
\endfirsthead
\caption[]{\textbf{(continued)} $R^2$ (in \%) of prediction of asymptotic stability by GNN and NetSciML in the WS-Kura experiment. GNN and NetSciML were trained on the degree $6$ and rewiring probability $0.5$ ensemble. HGBR (dyn. features) was trained using only the inhomogeneous nodal parameters as input ($P^d$, $D$ and total coupling strength at the node), HGBR (top. features) used only topological features.} \\
\toprule
 & \multicolumn{4}{c}{NetSciML} & \multicolumn{4}{c}{\makecell[l]{NetSciML \\(dyn. features)}} & \multicolumn{4}{c}{\makecell[l]{NetSciML\\(top. features)}} & \multicolumn{4}{c}{GNN} \\
 & 4 & 6 & 8 & 10 & 4 & 6 & 8 & 10 & 4 & 6 & 8 & 10 & 4 & 6 & 8 & 10 \\
\midrule
\endhead
\midrule
\multicolumn{17}{r}{Continued on next page} \\
\midrule
\endfoot
\bottomrule
\endlastfoot
0.1 & {\cellcolor[HTML]{F7FBFF}} \color[HTML]{000000} \color{gray} $-40$ & {\cellcolor[HTML]{9CC9E1}} \color[HTML]{000000} $38$ & {\cellcolor[HTML]{1562A9}} \color[HTML]{F1F1F1} $81$ & {\cellcolor[HTML]{125DA6}} \color[HTML]{F1F1F1} $83$ & {\cellcolor[HTML]{F7FBFF}} \color[HTML]{000000} \color{gray} $\ll 0$ & {\cellcolor[HTML]{BDD7EC}} \color[HTML]{000000} $28$ & {\cellcolor[HTML]{1967AD}} \color[HTML]{F1F1F1} $79$ & {\cellcolor[HTML]{1562A9}} \color[HTML]{F1F1F1} $81$ & {\cellcolor[HTML]{F7FBFF}} \color[HTML]{000000} \color{gray} $\ll 0$ & {\cellcolor[HTML]{C3DAEE}} \color[HTML]{000000} $26$ & {\cellcolor[HTML]{3A8AC2}} \color[HTML]{F1F1F1} $66$ & {\cellcolor[HTML]{58A1CF}} \color[HTML]{F1F1F1} $56$ & {\cellcolor[HTML]{F7FBFF}} \color[HTML]{000000} \color{gray} $-44$ & {\cellcolor[HTML]{77B5D9}} \color[HTML]{000000} $47$ & {\cellcolor[HTML]{2E7EBC}} \color[HTML]{F1F1F1} $70$ & {\cellcolor[HTML]{F7FBFF}} \color[HTML]{000000} \color{gray} $-7$ \\
0.3 & {\cellcolor[HTML]{E1EDF8}} \color[HTML]{000000} $11$ & {\cellcolor[HTML]{2C7CBA}} \color[HTML]{F1F1F1} $71$ & {\cellcolor[HTML]{084D96}} \color[HTML]{F1F1F1} $89$ & {\cellcolor[HTML]{1764AB}} \color[HTML]{F1F1F1} $80$ & {\cellcolor[HTML]{F7FBFF}} \color[HTML]{000000} \color{gray} $-65$ & {\cellcolor[HTML]{3686C0}} \color[HTML]{F1F1F1} $67$ & {\cellcolor[HTML]{0E59A2}} \color[HTML]{F1F1F1} $84$ & {\cellcolor[HTML]{1E6DB2}} \color[HTML]{F1F1F1} $76$ & {\cellcolor[HTML]{F7FBFF}} \color[HTML]{000000} \color{gray} $-37$ & {\cellcolor[HTML]{529DCC}} \color[HTML]{F1F1F1} $58$ & {\cellcolor[HTML]{2E7EBC}} \color[HTML]{F1F1F1} $70$ & {\cellcolor[HTML]{8ABFDD}} \color[HTML]{000000} $42$ & {\cellcolor[HTML]{C4DAEE}} \color[HTML]{000000} $25$ & {\cellcolor[HTML]{08509B}} \color[HTML]{F1F1F1} $88$ & {\cellcolor[HTML]{1663AA}} \color[HTML]{F1F1F1} $80$ & {\cellcolor[HTML]{F7FBFF}} \color[HTML]{000000} \color{gray} $-12$ \\
0.5 & {\cellcolor[HTML]{D6E5F4}} \color[HTML]{000000} $17$ & {\cellcolor[HTML]{1C6BB0}} \color[HTML]{F1F1F1} $77$ & {\cellcolor[HTML]{084F99}} \color[HTML]{F1F1F1} $88$ & {\cellcolor[HTML]{1C6BB0}} \color[HTML]{F1F1F1} $77$ & {\cellcolor[HTML]{F7FBFF}} \color[HTML]{000000} \color{gray} $-17$ & {\cellcolor[HTML]{2373B6}} \color[HTML]{F1F1F1} $74$ & {\cellcolor[HTML]{125DA6}} \color[HTML]{F1F1F1} $83$ & {\cellcolor[HTML]{2A7AB9}} \color[HTML]{F1F1F1} $72$ & {\cellcolor[HTML]{E9F2FA}} \color[HTML]{000000} $7$ & {\cellcolor[HTML]{4292C6}} \color[HTML]{F1F1F1} $62$ & {\cellcolor[HTML]{3888C1}} \color[HTML]{F1F1F1} $66$ & {\cellcolor[HTML]{75B4D8}} \color[HTML]{000000} $48$ & {\cellcolor[HTML]{4090C5}} \color[HTML]{F1F1F1} $63$ & {\cellcolor[HTML]{083B7C}} \color[HTML]{F1F1F1} $96$ & {\cellcolor[HTML]{1764AB}} \color[HTML]{F1F1F1} $80$ & {\cellcolor[HTML]{F7FBFF}} \color[HTML]{000000} \color{gray} $-12$ \\
0.7 & {\cellcolor[HTML]{DCEAF6}} \color[HTML]{000000} $14$ & {\cellcolor[HTML]{1E6DB2}} \color[HTML]{F1F1F1} $76$ & {\cellcolor[HTML]{08509B}} \color[HTML]{F1F1F1} $88$ & {\cellcolor[HTML]{1F6EB3}} \color[HTML]{F1F1F1} $76$ & {\cellcolor[HTML]{F7FBFF}} \color[HTML]{000000} \color{gray} $0$ & {\cellcolor[HTML]{2474B7}} \color[HTML]{F1F1F1} $74$ & {\cellcolor[HTML]{135FA7}} \color[HTML]{F1F1F1} $82$ & {\cellcolor[HTML]{2F7FBC}} \color[HTML]{F1F1F1} $70$ & {\cellcolor[HTML]{D0E2F2}} \color[HTML]{000000} $20$ & {\cellcolor[HTML]{4D99CA}} \color[HTML]{F1F1F1} $59$ & {\cellcolor[HTML]{3585BF}} \color[HTML]{F1F1F1} $68$ & {\cellcolor[HTML]{7CB7DA}} \color[HTML]{000000} $46$ & {\cellcolor[HTML]{3686C0}} \color[HTML]{F1F1F1} $67$ & {\cellcolor[HTML]{083B7C}} \color[HTML]{F1F1F1} $96$ & {\cellcolor[HTML]{1663AA}} \color[HTML]{F1F1F1} $80$ & {\cellcolor[HTML]{F7FBFF}} \color[HTML]{000000} \color{gray} $-15$ \\
0.9 & {\cellcolor[HTML]{DEEBF7}} \color[HTML]{000000} $13$ & {\cellcolor[HTML]{2A7AB9}} \color[HTML]{F1F1F1} $72$ & {\cellcolor[HTML]{09529D}} \color[HTML]{F1F1F1} $87$ & {\cellcolor[HTML]{2171B5}} \color[HTML]{F1F1F1} $75$ & {\cellcolor[HTML]{F0F6FD}} \color[HTML]{000000} $4$ & {\cellcolor[HTML]{2B7BBA}} \color[HTML]{F1F1F1} $71$ & {\cellcolor[HTML]{1663AA}} \color[HTML]{F1F1F1} $80$ & {\cellcolor[HTML]{3383BE}} \color[HTML]{F1F1F1} $68$ & {\cellcolor[HTML]{D0E2F2}} \color[HTML]{000000} $20$ & {\cellcolor[HTML]{60A7D2}} \color[HTML]{F1F1F1} $53$ & {\cellcolor[HTML]{3585BF}} \color[HTML]{F1F1F1} $67$ & {\cellcolor[HTML]{82BBDB}} \color[HTML]{000000} $44$ & {\cellcolor[HTML]{3D8DC4}} \color[HTML]{F1F1F1} $64$ & {\cellcolor[HTML]{083776}} \color[HTML]{F1F1F1} $97$ & {\cellcolor[HTML]{1663AA}} \color[HTML]{F1F1F1} $80$ & {\cellcolor[HTML]{F7FBFF}} \color[HTML]{000000} \color{gray} $-15$ \\
\end{longtable}

Graph neural networks (GNNs) trained at $(d=6,\, p_r=0.5)$ achieve moderate extrapolation to the sparse/regular regime, but their loss in accuracy mirrors that of NetSciML (\Cref{tab:WSKura NetSciML GNN}). GNNs outperform NetSciML in the dense regime, and part of this advantage carries over when extrapolating; however, models trained on dense data still fail to capture additional mechanisms driving (in)stability that emerge in sparse networks.

\subsection*{Training and model details} 
The GNN used is a DBGNN with 2 layers, each comprimising 20 internal steps. The hidden channel dimension for both nodes and edges is set to 500. No regression head is applied. The model has The model is trained for 2000 epochs with a batch size of 50. OneCycle learning rate scheduling is used with cosine anneal strategy, the final div factor is set to $9694991443.141111$, the initial div factor to 9363750, and a maximum learning rate of 180.978942.

\Cref{tab:WS-Kura-Transient_gnn_std} reports the mean and standard errors over multiple GNN initializations.
\begin{table}[htbp]
\centering
\caption{WS-Kura Transient: GNN $R^2$ scores reported as mean $\pm$ standard deviation across multiple random initializations.}
\label{tab:WS-Kura-Transient_gnn_std}
\begin{tabularx}{\linewidth}{l>{\centering\arraybackslash}X>{\centering\arraybackslash}X>{\centering\arraybackslash}X>{\centering\arraybackslash}X}
\toprule
 & 4 & 6 & 8 & 10 \\
\midrule
0.9 & {\cellcolor[HTML]{4292C6} \color[HTML]{F1F1F1} 0.644 $\pm$ 0.016} & {\cellcolor[HTML]{08306B} \color[HTML]{F1F1F1} 0.969 $\pm$ 0.001} & {\cellcolor[HTML]{2171B5} \color[HTML]{F1F1F1} 0.802 $\pm$ 0.030} & {\cellcolor[HTML]{F7FBFF} \color{gray} -0.145 $\pm$ 0.170} \\
0.7 & {\cellcolor[HTML]{4292C6} \color[HTML]{F1F1F1} 0.672 $\pm$ 0.011} & {\cellcolor[HTML]{08306B} \color[HTML]{F1F1F1} 0.957 $\pm$ 0.001} & {\cellcolor[HTML]{2171B5} \color[HTML]{F1F1F1} 0.802 $\pm$ 0.031} & {\cellcolor[HTML]{F7FBFF} \color{gray} -0.154 $\pm$ 0.175} \\
0.5 & {\cellcolor[HTML]{4292C6} \color[HTML]{F1F1F1} 0.632 $\pm$ 0.014} & {\cellcolor[HTML]{08306B} \color[HTML]{F1F1F1} 0.957 $\pm$ 0.003} & {\cellcolor[HTML]{2171B5} \color[HTML]{F1F1F1} 0.799 $\pm$ 0.032} & {\cellcolor[HTML]{F7FBFF} \color{gray} -0.124 $\pm$ 0.174} \\
0.3 & {\cellcolor[HTML]{C6DBEF} \color[HTML]{000000} 0.254 $\pm$ 0.033} & {\cellcolor[HTML]{08519C} \color[HTML]{F1F1F1} 0.877 $\pm$ 0.003} & {\cellcolor[HTML]{2171B5} \color[HTML]{F1F1F1} 0.802 $\pm$ 0.032} & {\cellcolor[HTML]{F7FBFF} \color{gray} -0.117 $\pm$ 0.180} \\
0.1 & {\cellcolor[HTML]{F7FBFF} \color{gray} -0.445 $\pm$ 0.065} & {\cellcolor[HTML]{6BAED6} \color[HTML]{000000} 0.471 $\pm$ 0.013} & {\cellcolor[HTML]{2171B5} \color[HTML]{F1F1F1} 0.702 $\pm$ 0.036} & {\cellcolor[HTML]{F7FBFF} \color{gray} -0.070 $\pm$ 0.179} \\
\bottomrule
\end{tabularx}
\end{table}

\FloatBarrier
\section{Supplementary material on WS-VDP experiments}
\label{sec:appendix_experiments_WSVDP}
We consider Watts–Strogatz ensembles with mean degree $d \in \{4,6,8,10\}$ and rewiring probability $p_r \in \{0.1,0.3,0.5,0.7,0.9\}$. Node dynamics are van der Pol oscillators with homogeneous parameters chosen to yield a bistable regime. Asymptotic stability is assessed via single-node basin stability (return of frequency to narrow bounds), and transient stability via a synchronization norm.

\begin{longtable}{p{3cm}cccccccccc}
\caption{Pearson correlation (in \%) of network measures and stability index in the WS-VDP experiment.} \label{tab:VDP all correlations} \\
\toprule
 & \multicolumn{5}{c}{Asymptotic} & \multicolumn{5}{c}{Transient} \\
 & 6 0.5 & 4 0.3 & 8 0.3 & 4 0.7 & 8 0.7 & 6 0.5 & 4 0.3 & 8 0.3 & 4 0.7 & 8 0.7 \\
Network Measure &  &  &  &  &  &  &  &  &  &  \\
\midrule
\endfirsthead
\caption[]{\textbf{(continued)} Pearson correlation (in \%) of network measures and stability index in the WS-VDP experiment.} \\
\toprule
 & \multicolumn{5}{c}{Asymptotic} & \multicolumn{5}{c}{Transient} \\
 & 6 0.5 & 4 0.3 & 8 0.3 & 4 0.7 & 8 0.7 & 6 0.5 & 4 0.3 & 8 0.3 & 4 0.7 & 8 0.7 \\
Network Measure &  &  &  &  &  &  &  &  &  &  \\
\midrule
\endhead
\midrule
\multicolumn{11}{r}{Continued on next page} \\
\midrule
\endfoot
\bottomrule
\endlastfoot
\textsf{current-flow closeness cent.} & {\cellcolor[HTML]{F7AC52}} \color[HTML]{000000} $-35$ & {\cellcolor[HTML]{FBEAD2}} \color[HTML]{000000} $-9$ & {\cellcolor[HTML]{9D94C2}} \color[HTML]{F1F1F1} $38$ & {\cellcolor[HTML]{FED095}} \color[HTML]{000000} $-22$ & {\cellcolor[HTML]{C3C0DD}} \color[HTML]{000000} $25$ & {\cellcolor[HTML]{F8F3EC}} \color[HTML]{000000} $-3$ & {\cellcolor[HTML]{EF9E3C}} \color[HTML]{000000} $-39$ & {\cellcolor[HTML]{DB7D12}} \color[HTML]{F1F1F1} $-50$ & {\cellcolor[HTML]{FBB55E}} \color[HTML]{000000} $-33$ & {\cellcolor[HTML]{7F3B08}} \color[HTML]{F1F1F1} $-84$ \\
\cline{1-11}
\textsf{resistance distance cent.} & {\cellcolor[HTML]{F7AC52}} \color[HTML]{000000} $-35$ & {\cellcolor[HTML]{FBEAD2}} \color[HTML]{000000} $-9$ & {\cellcolor[HTML]{9D94C2}} \color[HTML]{F1F1F1} $38$ & {\cellcolor[HTML]{FED095}} \color[HTML]{000000} $-22$ & {\cellcolor[HTML]{C3C0DD}} \color[HTML]{000000} $25$ & {\cellcolor[HTML]{F8F3EC}} \color[HTML]{000000} $-3$ & {\cellcolor[HTML]{EF9E3C}} \color[HTML]{000000} $-39$ & {\cellcolor[HTML]{DB7D12}} \color[HTML]{F1F1F1} $-50$ & {\cellcolor[HTML]{FBB55E}} \color[HTML]{000000} $-33$ & {\cellcolor[HTML]{7F3B08}} \color[HTML]{F1F1F1} $-84$ \\
\cline{1-11}
\textsf{harmonic cent.} & {\cellcolor[HTML]{FBB55E}} \color[HTML]{000000} $-33$ & {\cellcolor[HTML]{FAECD7}} \color[HTML]{000000} $-8$ & {\cellcolor[HTML]{968BBC}} \color[HTML]{F1F1F1} $41$ & {\cellcolor[HTML]{FDCC8C}} \color[HTML]{000000} $-24$ & {\cellcolor[HTML]{C9C8E1}} \color[HTML]{000000} $22$ & {\cellcolor[HTML]{FBEAD2}} \color[HTML]{000000} $-9$ & {\cellcolor[HTML]{EE9B39}} \color[HTML]{000000} $-40$ & {\cellcolor[HTML]{C76B0C}} \color[HTML]{F1F1F1} $-57$ & {\cellcolor[HTML]{FCB761}} \color[HTML]{000000} $-32$ & {\cellcolor[HTML]{7F3B08}} \color[HTML]{F1F1F1} $-83$ \\
\end{longtable}

\begin{longtable}{p{3cm}cccccccccc}
\caption{Theil's U (in \%) of network measures and stability index in the WS-VDP experiment.} \label{tab:VDP all TU} \\
\toprule
 & \multicolumn{5}{c}{Asymptotic} & \multicolumn{5}{c}{Transient} \\
 & 6 0.5 & 4 0.3 & 8 0.3 & 4 0.7 & 8 0.7 & 6 0.5 & 4 0.3 & 8 0.3 & 4 0.7 & 8 0.7 \\
Network Measure &  &  &  &  &  &  &  &  &  &  \\
\midrule
\endfirsthead
\caption[]{\textbf{(continued)} Theil's U (in \%) of network measures and stability index in the WS-VDP experiment.} \\
\toprule
 & \multicolumn{5}{c}{Asymptotic} & \multicolumn{5}{c}{Transient} \\
 & 6 0.5 & 4 0.3 & 8 0.3 & 4 0.7 & 8 0.7 & 6 0.5 & 4 0.3 & 8 0.3 & 4 0.7 & 8 0.7 \\
Network Measure &  &  &  &  &  &  &  &  &  &  \\
\midrule
\endhead
\midrule
\multicolumn{11}{r}{Continued on next page} \\
\midrule
\endfoot
\bottomrule
\endlastfoot
\textsf{is bulk} & {\cellcolor[HTML]{F6F6F7}} \color[HTML]{000000} $0$ & {\cellcolor[HTML]{F6F6F7}} \color[HTML]{000000} $0$ & {\cellcolor[HTML]{F6F6F7}} \color[HTML]{000000} $0$ & {\cellcolor[HTML]{F6F6F7}} \color[HTML]{000000} $0$ & {\cellcolor[HTML]{F6F6F7}} \color[HTML]{000000} $0$ & {\cellcolor[HTML]{F6F6F7}} \color[HTML]{000000} $0$ & {\cellcolor[HTML]{F6F6F7}} \color[HTML]{000000} $0$ & {\cellcolor[HTML]{F6F6F7}} \color[HTML]{000000} $0$ & {\cellcolor[HTML]{F6F6F7}} \color[HTML]{000000} $0$ & {\cellcolor[HTML]{F6F6F7}} \color[HTML]{000000} $0$ \\
\cline{1-11}
\textsf{is dense sprout} & {\cellcolor[HTML]{F6F6F7}} \color[HTML]{000000} $0$ & {\cellcolor[HTML]{F6F6F7}} \color[HTML]{000000} $0$ & {\cellcolor[HTML]{F6F6F7}} \color[HTML]{000000} $0$ & {\cellcolor[HTML]{F6F6F7}} \color[HTML]{000000} $0$ & {\cellcolor[HTML]{F6F6F7}} \color[HTML]{000000} $0$ & {\cellcolor[HTML]{F6F6F7}} \color[HTML]{000000} $0$ & {\cellcolor[HTML]{F6F6F7}} \color[HTML]{000000} $0$ & {\cellcolor[HTML]{F6F6F7}} \color[HTML]{000000} $0$ & {\cellcolor[HTML]{F6F6F7}} \color[HTML]{000000} $0$ & {\cellcolor[HTML]{F6F6F7}} \color[HTML]{000000} $0$ \\
\cline{1-11}
\textsf{proper leaf node} & {\cellcolor[HTML]{F6F6F7}} \color[HTML]{000000} $0$ & {\cellcolor[HTML]{F6F6F7}} \color[HTML]{000000} $0$ & {\cellcolor[HTML]{F6F6F7}} \color[HTML]{000000} $0$ & {\cellcolor[HTML]{F6F6F7}} \color[HTML]{000000} $0$ & {\cellcolor[HTML]{F6F6F7}} \color[HTML]{000000} $0$ & {\cellcolor[HTML]{F6F6F7}} \color[HTML]{000000} $0$ & {\cellcolor[HTML]{F6F6F7}} \color[HTML]{000000} $0$ & {\cellcolor[HTML]{F6F6F7}} \color[HTML]{000000} $0$ & {\cellcolor[HTML]{F6F6F7}} \color[HTML]{000000} $0$ & {\cellcolor[HTML]{F6F6F7}} \color[HTML]{000000} $0$ \\
\end{longtable}

\begin{longtable}{p{3cm}rrrrrrrrr}
\caption{Theil's U (in \%) of selected network measures and stability index the WS-VDP experiment.} \label{tab:VDP tu selected nm} \\
\toprule
 &  & \multicolumn{4}{c}{Transient} & \multicolumn{4}{c}{Asymptotic} \\
 &  & 4 & 6 & 8 & 10 & 4 & 6 & 8 & 10 \\
Network Measure &  &  &  &  &  &  &  &  &  \\
\midrule
\endfirsthead
\caption[]{\textbf{(continued)} Theil's U (in \%) of selected network measures and stability index the WS-VDP experiment.} \\
\toprule
 &  & \multicolumn{4}{c}{Transient} & \multicolumn{4}{c}{Asymptotic} \\
 &  & 4 & 6 & 8 & 10 & 4 & 6 & 8 & 10 \\
Network Measure &  &  &  &  &  &  &  &  &  \\
\midrule
\endhead
\midrule
\multicolumn{10}{r}{Continued on next page} \\
\midrule
\endfoot
\bottomrule
\endlastfoot
\multirow[c]{5}{=}{\textsf{node deg.}} & 0.1 & {\cellcolor[HTML]{F0F1F4}} \color[HTML]{000000} $3$ & {\cellcolor[HTML]{F2F2F5}} \color[HTML]{000000} $3$ & {\cellcolor[HTML]{F0F1F4}} \color[HTML]{000000} $4$ & {\cellcolor[HTML]{D2D3E7}} \color[HTML]{000000} $18$ & {\cellcolor[HTML]{EBECF3}} \color[HTML]{000000} $6$ & {\cellcolor[HTML]{E9EAF2}} \color[HTML]{000000} $7$ & {\cellcolor[HTML]{EAEBF2}} \color[HTML]{000000} $7$ & {\cellcolor[HTML]{CFCFE5}} \color[HTML]{000000} $20$ \\*
 & 0.3 & {\cellcolor[HTML]{E9EAF2}} \color[HTML]{000000} $7$ & {\cellcolor[HTML]{EBECF3}} \color[HTML]{000000} $6$ & {\cellcolor[HTML]{D4D4E8}} \color[HTML]{000000} $18$ & {\cellcolor[HTML]{C8C6E0}} \color[HTML]{000000} $23$ & {\cellcolor[HTML]{DFE1EE}} \color[HTML]{000000} $12$ & {\cellcolor[HTML]{D4D4E8}} \color[HTML]{000000} $18$ & {\cellcolor[HTML]{C0BDDB}} \color[HTML]{000000} $26$ & {\cellcolor[HTML]{BFBBDA}} \color[HTML]{000000} $27$ \\*
 & 0.5 & {\cellcolor[HTML]{E5E7F0}} \color[HTML]{000000} $9$ & {\cellcolor[HTML]{D7D8EA}} \color[HTML]{000000} $17$ & {\cellcolor[HTML]{A59DC8}} \color[HTML]{F1F1F1} $36$ & {\cellcolor[HTML]{968BBC}} \color[HTML]{F1F1F1} $41$ & {\cellcolor[HTML]{D7D8EA}} \color[HTML]{000000} $17$ & {\cellcolor[HTML]{AFA8D0}} \color[HTML]{000000} $33$ & {\cellcolor[HTML]{7D6DA9}} \color[HTML]{F1F1F1} $49$ & {\cellcolor[HTML]{7D6DA9}} \color[HTML]{F1F1F1} $49$ \\*
 & 0.7 & {\cellcolor[HTML]{E5E7F0}} \color[HTML]{000000} $9$ & {\cellcolor[HTML]{C9C8E1}} \color[HTML]{000000} $22$ & {\cellcolor[HTML]{9287B9}} \color[HTML]{F1F1F1} $42$ & {\cellcolor[HTML]{8E82B6}} \color[HTML]{F1F1F1} $43$ & {\cellcolor[HTML]{D4D4E8}} \color[HTML]{000000} $18$ & {\cellcolor[HTML]{9D94C2}} \color[HTML]{F1F1F1} $39$ & {\cellcolor[HTML]{5D3790}} \color[HTML]{F1F1F1} $60$ & {\cellcolor[HTML]{8C80B5}} \color[HTML]{F1F1F1} $44$ \\
\end{longtable}

\Cref{tab:VDP all correlations} reports Pearson correlations for the network measures in WS-Kura, while \Cref{tab:VDP all TU} reports Theil’s $U$. For a selected subset of network measures, \Cref{tab:VDP pearson selected nm} provides the corresponding Pearson correlations.

For the VDP dynamics, correlations between network structure and stability vary markedly across ensembles \Cref{tab:VDP pearson selected nm}. For instance, the correlation of resistance distance centrality with transient stability is $+0.08$ for $d=6$, $p_r=0.3$ but $-0.8$ for $d=8$, $p_r=0.5$, indicating pronounced sign changes and limited consistency across regimes.

Models trained at $(d=6,\, p_r=0.5)$ generalize unreliably: neither GNNs (\Cref{tab:VDP NetSciML}) nor NetSciML transfer well beyond increases in rewiring probability. Overall, the GNN underperforms NetSciML even in-distribution, and both approaches fail to capture stability mechanisms that shift with topology in sparser or more regular networks.

\subsection*{Training and model details}
The GNN used is a DBGNN with 2 layers, each comprising 10 internal steps. The hidden channel dimension for both nodes and edges is set to 100. A linear layer of dimension 100 follows the GNN convolutions. The model consists of 111,001 parameters. Dropout is set to 0. The batch size is 32, the learning rate is $1 \times 10^{-5}$, and the model is trained for 2000 epochs. Training on 5 seeds consecutively with an H100 GPU takes approximately 15 hours.

\Cref{tab:WS-VDP-Asymptotic_gnn_std} reports the mean and standard errors over different initializations of the ML model.
\begin{table}[htbp]
\centering
\caption{WS-VDP Asymptotic: GNN $R^2$ scores reported as mean $\pm$ standard deviation across multiple random initializations.}
\label{tab:WS-VDP-Asymptotic_gnn_std}
\begin{tabularx}{\linewidth}{l>{\centering\arraybackslash}X>{\centering\arraybackslash}X>{\centering\arraybackslash}X>{\centering\arraybackslash}X}
\toprule
 & 4 & 6 & 8 & 10 \\
\midrule
0.9 & {\cellcolor[HTML]{F7FBFF} \color{gray} -8.700 $\pm$ 0.495} & {\cellcolor[HTML]{4292C6} \color[HTML]{F1F1F1} 0.653 $\pm$ 0.006} & {\cellcolor[HTML]{F7FBFF} \color{gray} -21.325 $\pm$ 4.522} & {\cellcolor[HTML]{F7FBFF} \color{gray} -308.414 $\pm$ 83.037} \\
0.7 & {\cellcolor[HTML]{F7FBFF} \color{gray} -8.460 $\pm$ 0.473} & {\cellcolor[HTML]{4292C6} \color[HTML]{F1F1F1} 0.645 $\pm$ 0.005} & {\cellcolor[HTML]{F7FBFF} \color{gray} -21.592 $\pm$ 4.657} & {\cellcolor[HTML]{F7FBFF} \color{gray} -313.146 $\pm$ 86.046} \\
0.5 & {\cellcolor[HTML]{F7FBFF} \color{gray} -7.722 $\pm$ 0.456} & {\cellcolor[HTML]{6BAED6} \color[HTML]{000000} 0.549 $\pm$ 0.005} & {\cellcolor[HTML]{F7FBFF} \color{gray} -19.425 $\pm$ 4.281} & {\cellcolor[HTML]{F7FBFF} \color{gray} -318.296 $\pm$ 90.731} \\
0.3 & {\cellcolor[HTML]{F7FBFF} \color{gray} -7.741 $\pm$ 0.576} & {\cellcolor[HTML]{DEEBF7} \color[HTML]{000000} 0.079 $\pm$ 0.018} & {\cellcolor[HTML]{F7FBFF} \color{gray} -7.806 $\pm$ 1.996} & {\cellcolor[HTML]{F7FBFF} \color{gray} -91.260 $\pm$ 28.640} \\
0.1 & {\cellcolor[HTML]{F7FBFF} \color{gray} -8.995 $\pm$ 0.704} & {\cellcolor[HTML]{F7FBFF} \color{gray} -0.917 $\pm$ 0.027} & {\cellcolor[HTML]{F7FBFF} \color{gray} -0.083 $\pm$ 0.076} & {\cellcolor[HTML]{F7FBFF} \color{gray} -7.093 $\pm$ 3.132} \\
\bottomrule
\end{tabularx}
\end{table}

\FloatBarrier

\section{Supplementary material on PG-Kura experiments}
\label{sec:appendix_experiments_PGKura}
We analyze synthetic grids with 20 and 100 nodes, a large synthetic Texas system (approximately 2000 nodes) generated via a distinct procedure, and several real-world topologies. Node dynamics follow an inertial Kuramoto model with nearly homogeneous parameters; the only node-specific variation is the injected power $P^d$. Asymptotic stability is assessed via single-node basin stability (return of frequency to narrow bounds), and transient stability via the maximum frequency deviation.

\begin{longtable}{p{2.5cm}p{3cm}rrrrrrr}
\caption{Pearson correlation (in \%) of network measures and stability index in the PG-Kura experiment for network measures with highest absolute correlation across grid sizes.} \label{tab:PGKura abs correlations} \\
\toprule
 &  & 20 & 100 & DE & FR & GB & ES & TX \\
Stability & Network Measure &  &  &  &  &  &  &  \\
\midrule
\endfirsthead
\caption[]{\textbf{(continued)} Pearson correlation (in \%) of network measures and stability index in the PG-Kura experiment for network measures with highest absolute correlation across grid sizes.} \\
\toprule
 &  & 20 & 100 & DE & FR & GB & ES & TX \\
Stability & Network Measure &  &  &  &  &  &  &  \\
\midrule
\endhead
\midrule
\multicolumn{9}{r}{Continued on next page} \\
\midrule
\endfoot
\bottomrule
\endlastfoot
\multirow[c]{4}{=}{Asymptotic} & \textsf{second-order cent.} & {\cellcolor[HTML]{B65A07}} \color[HTML]{F1F1F1} $-63$ & {\cellcolor[HTML]{AA5306}} \color[HTML]{F1F1F1} $-67$ & {\cellcolor[HTML]{C0640A}} \color[HTML]{F1F1F1} $-59$ & {\cellcolor[HTML]{A85206}} \color[HTML]{F1F1F1} $-67$ & {\cellcolor[HTML]{FCB761}} \color[HTML]{000000} $-32$ & {\cellcolor[HTML]{813C08}} \color[HTML]{F1F1F1} $-79$ & {\cellcolor[HTML]{B95E08}} \color[HTML]{F1F1F1} $-61$ \\*
 & \textsf{resistance distance cent.} & {\cellcolor[HTML]{502382}} \color[HTML]{F1F1F1} $65$ & {\cellcolor[HTML]{502382}} \color[HTML]{F1F1F1} $65$ & {\cellcolor[HTML]{664697}} \color[HTML]{F1F1F1} $57$ & {\cellcolor[HTML]{4A1D78}} \color[HTML]{F1F1F1} $68$ & {\cellcolor[HTML]{ADA6CE}} \color[HTML]{F1F1F1} $33$ & {\cellcolor[HTML]{360959}} \color[HTML]{F1F1F1} $76$ & {\cellcolor[HTML]{532687}} \color[HTML]{F1F1F1} $64$ \\*
 & \textsf{current-flow closeness cent.} & {\cellcolor[HTML]{502382}} \color[HTML]{F1F1F1} $65$ & {\cellcolor[HTML]{502382}} \color[HTML]{F1F1F1} $65$ & {\cellcolor[HTML]{664697}} \color[HTML]{F1F1F1} $57$ & {\cellcolor[HTML]{4A1D78}} \color[HTML]{F1F1F1} $68$ & {\cellcolor[HTML]{ADA6CE}} \color[HTML]{F1F1F1} $33$ & {\cellcolor[HTML]{360959}} \color[HTML]{F1F1F1} $76$ & {\cellcolor[HTML]{532687}} \color[HTML]{F1F1F1} $64$ \\*
 & \textsf{harmonic cent.} & {\cellcolor[HTML]{572B8A}} \color[HTML]{F1F1F1} $63$ & {\cellcolor[HTML]{634094}} \color[HTML]{F1F1F1} $58$ & {\cellcolor[HTML]{968BBC}} \color[HTML]{F1F1F1} $41$ & {\cellcolor[HTML]{644395}} \color[HTML]{F1F1F1} $58$ & {\cellcolor[HTML]{9489BB}} \color[HTML]{F1F1F1} $41$ & {\cellcolor[HTML]{471A74}} \color[HTML]{F1F1F1} $69$ & {\cellcolor[HTML]{B4AED3}} \color[HTML]{000000} $31$ \\
\end{longtable}

\begin{longtable}{p{2.5cm}p{3cm}rrrrrrr}
\caption{Pearson correlation (in \%) of network measures and stability index in the PG-Kura experiment for network measures with highest variance of correlation across grid sizes.} \label{tab:PGKura var correlations} \\
\toprule
 &  & 20 & 100 & DE & FR & GB & ES & TX \\
Stability & Network Measure &  &  &  &  &  &  &  \\
\midrule
\endfirsthead
\caption[]{\textbf{(continued)} Pearson correlation (in \%) of network measures and stability index in the PG-Kura experiment for network measures with highest variance of correlation across grid sizes.} \\
\toprule
 &  & 20 & 100 & DE & FR & GB & ES & TX \\
Stability & Network Measure &  &  &  &  &  &  &  \\
\midrule
\endhead
\midrule
\multicolumn{9}{r}{Continued on next page} \\
\midrule
\endfoot
\bottomrule
\endlastfoot
\multirow[c]{4}{=}{Asymptotic} & \textsf{average neighbor deg.} & {\cellcolor[HTML]{CCCBE3}} \color[HTML]{000000} $21$ & {\cellcolor[HTML]{B3ACD2}} \color[HTML]{000000} $32$ & {\cellcolor[HTML]{F8F5F1}} \color[HTML]{000000} $-2$ & {\cellcolor[HTML]{C6C4DF}} \color[HTML]{000000} $24$ & {\cellcolor[HTML]{CCCBE3}} \color[HTML]{000000} $21$ & {\cellcolor[HTML]{8A7EB3}} \color[HTML]{F1F1F1} $45$ & {\cellcolor[HTML]{D57811}} \color[HTML]{F1F1F1} $-52$ \\*
 & \textsf{maximum neighbor deg.} & {\cellcolor[HTML]{8A7EB3}} \color[HTML]{F1F1F1} $45$ & {\cellcolor[HTML]{7764A5}} \color[HTML]{F1F1F1} $51$ & {\cellcolor[HTML]{C0BDDB}} \color[HTML]{000000} $26$ & {\cellcolor[HTML]{968BBC}} \color[HTML]{F1F1F1} $41$ & {\cellcolor[HTML]{BDB9D9}} \color[HTML]{000000} $27$ & {\cellcolor[HTML]{522584}} \color[HTML]{F1F1F1} $65$ & {\cellcolor[HTML]{FECF92}} \color[HTML]{000000} $-23$ \\*
 & \textsf{minimum neighbor deg.} & {\cellcolor[HTML]{FCE6C8}} \color[HTML]{000000} $-12$ & {\cellcolor[HTML]{FAEDDA}} \color[HTML]{000000} $-7$ & {\cellcolor[HTML]{FDC782}} \color[HTML]{000000} $-26$ & {\cellcolor[HTML]{FBEBD5}} \color[HTML]{000000} $-9$ & {\cellcolor[HTML]{E7E8F1}} \color[HTML]{000000} $9$ & {\cellcolor[HTML]{F8F4EE}} \color[HTML]{000000} $-2$ & {\cellcolor[HTML]{AA5306}} \color[HTML]{F1F1F1} $-67$ \\*
 & \textsf{eigenvector cent.} & {\cellcolor[HTML]{644395}} \color[HTML]{F1F1F1} $58$ & {\cellcolor[HTML]{8C80B5}} \color[HTML]{F1F1F1} $44$ & {\cellcolor[HTML]{DDDFED}} \color[HTML]{000000} $13$ & {\cellcolor[HTML]{A198C5}} \color[HTML]{F1F1F1} $37$ & {\cellcolor[HTML]{A198C5}} \color[HTML]{F1F1F1} $37$ & {\cellcolor[HTML]{CECDE4}} \color[HTML]{000000} $20$ & {\cellcolor[HTML]{F8F4EE}} \color[HTML]{000000} $-2$ \\
\end{longtable}

\begin{longtable}{p{2.5cm}p{2.5cm}rrrrrrr}
\caption{$R^2$ (in \%) of prediction of stability by NetSciML and GNNs in the PG-Kura experiment. Both NetSciML and GNN were trained on the 100 node ensemble.} \label{tab:PGKura hgbr gnn} \\
\toprule
 &  & 20 & 100 & DE & FR & GB & ES & TX \\
Stability & Prediction &  &  &  &  &  &  &  \\
\midrule
\endfirsthead
\caption[]{\textbf{(continued)} $R^2$ (in \%) of prediction of stability by NetSciML and GNNs in the PG-Kura experiment. Both NetSciML and GNN were trained on the 100 node ensemble.} \\
\toprule
 &  & 20 & 100 & DE & FR & GB & ES & TX \\
Stability & Prediction &  &  &  &  &  &  &  \\
\midrule
\endhead
\midrule
\multicolumn{9}{r}{Continued on next page} \\
\midrule
\endfoot
\bottomrule
\endlastfoot
\multirow[c]{2}{=}{Asymptotic} & GNN & {\cellcolor[HTML]{2A7AB9}} \color[HTML]{F1F1F1} $72$ & {\cellcolor[HTML]{084E98}} \color[HTML]{F1F1F1} $88$ & {\cellcolor[HTML]{08478D}} \color[HTML]{F1F1F1} $91$ & {\cellcolor[HTML]{08488E}} \color[HTML]{F1F1F1} $91$ & {\cellcolor[HTML]{4896C8}} \color[HTML]{F1F1F1} $61$ & {\cellcolor[HTML]{0F5AA3}} \color[HTML]{F1F1F1} $84$ & {\cellcolor[HTML]{105BA4}} \color[HTML]{F1F1F1} $83$ \\*
 & NetSciML & {\cellcolor[HTML]{9FCAE1}} \color[HTML]{000000} $37$ & {\cellcolor[HTML]{125EA6}} \color[HTML]{F1F1F1} $82$ & {\cellcolor[HTML]{2F7FBC}} \color[HTML]{F1F1F1} $70$ & {\cellcolor[HTML]{2777B8}} \color[HTML]{F1F1F1} $72$ & {\cellcolor[HTML]{B7D4EA}} \color[HTML]{000000} $30$ & {\cellcolor[HTML]{1A68AE}} \color[HTML]{F1F1F1} $78$ & {\cellcolor[HTML]{4A98C9}} \color[HTML]{F1F1F1} $60$ \\
\cline{1-9}
\multirow[c]{2}{=}{Transient} & GNN & {\cellcolor[HTML]{083B7C}} \color[HTML]{F1F1F1} $96$ & {\cellcolor[HTML]{083573}} \color[HTML]{F1F1F1} $98$ & {\cellcolor[HTML]{08306B}} \color[HTML]{F1F1F1} $100$ & {\cellcolor[HTML]{08316D}} \color[HTML]{F1F1F1} $99$ & {\cellcolor[HTML]{08316D}} \color[HTML]{F1F1F1} $99$ & {\cellcolor[HTML]{08316D}} \color[HTML]{F1F1F1} $100$ & {\cellcolor[HTML]{083B7C}} \color[HTML]{F1F1F1} $96$ \\*
 & NetSciML & {\cellcolor[HTML]{08478D}} \color[HTML]{F1F1F1} $91$ & {\cellcolor[HTML]{083674}} \color[HTML]{F1F1F1} $97$ & {\cellcolor[HTML]{083776}} \color[HTML]{F1F1F1} $97$ & {\cellcolor[HTML]{083A7A}} \color[HTML]{F1F1F1} $96$ & {\cellcolor[HTML]{08316D}} \color[HTML]{F1F1F1} $99$ & {\cellcolor[HTML]{08306B}} \color[HTML]{F1F1F1} $100$ & {\cellcolor[HTML]{1C6AB0}} \color[HTML]{F1F1F1} $78$ \\
\end{longtable}

\begin{longtable}{p{2.5cm}p{3.5cm}rrrrrrr}
\caption{$R^2$ score (in \%) of NetSciML trained with reduced data on the 20 and 100 node ensembles in the PG-Kura experiment. HGBR-20/100, N indicates the HGBR model trained on N networks from the 20/100 node ensemble, 'no PF' indicates the HGBR model trained without the features derived from the synchronous state, OMP3 is linear regression using orthogonal matching pursuit to select the 3 most important network measures.} \label{tab:PGKura hgbr data efficiency} \\
\toprule
 & Nodes & 20 & 100 & DE & FR & GB & ES & TX \\
Stability & NetSciML &  &  &  &  &  &  &  \\
\midrule
\endfirsthead
\caption[]{\textbf{(continued)} $R^2$ score (in \%) of NetSciML trained with reduced data on the 20 and 100 node ensembles in the PG-Kura experiment. HGBR-20/100, N indicates the HGBR model trained on N networks from the 20/100 node ensemble, 'no PF' indicates the HGBR model trained without the features derived from the synchronous state, OMP3 is linear regression using orthogonal matching pursuit to select the 3 most important network measures.} \\
\toprule
 & Nodes & 20 & 100 & DE & FR & GB & ES & TX \\
Stability & NetSciML &  &  &  &  &  &  &  \\
\midrule
\endhead
\midrule
\multicolumn{9}{r}{Continued on next page} \\
\midrule
\endfoot
\bottomrule
\endlastfoot
\multirow[c]{11}{=}{Transient} & HGBR-100 & {\cellcolor[HTML]{08478D}} \color[HTML]{F1F1F1} $91$ & {\cellcolor[HTML]{083674}} \color[HTML]{F1F1F1} $97$ & {\cellcolor[HTML]{083776}} \color[HTML]{F1F1F1} $97$ & {\cellcolor[HTML]{083A7A}} \color[HTML]{F1F1F1} $96$ & {\cellcolor[HTML]{08316D}} \color[HTML]{F1F1F1} $99$ & {\cellcolor[HTML]{08306B}} \color[HTML]{F1F1F1} $100$ & {\cellcolor[HTML]{1C6AB0}} \color[HTML]{F1F1F1} $78$ \\*
 & HGBR-100, 7 & {\cellcolor[HTML]{084C95}} \color[HTML]{F1F1F1} $89$ & {\cellcolor[HTML]{084285}} \color[HTML]{F1F1F1} $93$ & {\cellcolor[HTML]{084B93}} \color[HTML]{F1F1F1} $89$ & {\cellcolor[HTML]{08458A}} \color[HTML]{F1F1F1} $92$ & {\cellcolor[HTML]{084D96}} \color[HTML]{F1F1F1} $89$ & {\cellcolor[HTML]{08326E}} \color[HTML]{F1F1F1} $99$ & {\cellcolor[HTML]{1C6AB0}} \color[HTML]{F1F1F1} $77$ \\*
 & HGBR-100, 70 & {\cellcolor[HTML]{084488}} \color[HTML]{F1F1F1} $92$ & {\cellcolor[HTML]{083979}} \color[HTML]{F1F1F1} $96$ & {\cellcolor[HTML]{08458A}} \color[HTML]{F1F1F1} $92$ & {\cellcolor[HTML]{084082}} \color[HTML]{F1F1F1} $94$ & {\cellcolor[HTML]{083A7A}} \color[HTML]{F1F1F1} $96$ & {\cellcolor[HTML]{08306B}} \color[HTML]{F1F1F1} $100$ & {\cellcolor[HTML]{2070B4}} \color[HTML]{F1F1F1} $75$ \\*
 & HGBR-100, 700 & {\cellcolor[HTML]{084387}} \color[HTML]{F1F1F1} $93$ & {\cellcolor[HTML]{083776}} \color[HTML]{F1F1F1} $97$ & {\cellcolor[HTML]{084184}} \color[HTML]{F1F1F1} $94$ & {\cellcolor[HTML]{083C7D}} \color[HTML]{F1F1F1} $95$ & {\cellcolor[HTML]{08326E}} \color[HTML]{F1F1F1} $99$ & {\cellcolor[HTML]{08306B}} \color[HTML]{F1F1F1} $100$ & {\cellcolor[HTML]{2070B4}} \color[HTML]{F1F1F1} $75$ \\
\end{longtable}%

Correlations between structural descriptors and stability vary substantially across systems. For example, the correlation between average neighbor degree and asymptotic stability is $0.45$ in Spain but $-0.52$ in Texas, whereas second-order centrality and resistance-distance centrality remain comparatively stable (\Cref{tab:PGKura var correlations}). \Cref{tab:PGKura abs correlations} reports the most strongly correlated network measures, and \Cref{tab:PGKura var correlations} the network measures with the highest variance. 

Transient stability is predicted well by NetSciML \Cref{tab:PGKura hgbr gnn}. For asymptotic stability, GNNs trained on the 100-node ensemble generalize robustly to real-world grids (\Cref{tab:PGKura hgbr gnn}). A NetSciML model without synchronous-state (here also the power-flow) features attains comparable accuracy within the 100-node ensemble but fails to generalize to real-world topologies; NetSciML augmented with the full feature set improves generalization yet still lags behind GNNs.

These results indicate that the power-flow structure is a key determinant of stability. Network measures combined with the single dynamical feature $P^d$ do not recover this information. Even with power-flow features, NetSciML generalizes less robustly than GNNs: performance degrades on both smaller (20-node) and much larger (Texas) systems, whereas GNN accuracy remains largely stable across sizes, with the notable exception of GB.

\Cref{tab:PGKura hgbr data efficiency} reports the performance with fewer training samples. This table includes OMP3, which selects the subset of three network measures that collectively delivers the best predictive performance (see \Cref{sec:app_best_subset_selection} for details).

\subsection*{Details on the study of power grids with Kuramoto oscillators}
\label{sec:supplement_kuramoto_powergrids}

As a key challenge, we analyze the capability of predicting the dynamic stability of larger grids generated by the same random process, an even larger grid generated by a different random process and real-world topologies. 

This can be interpreted as measuring how well the prediction captures underlying causal relationships, rather than simply fitting statistical relationships specific to the ensemble. In power grid applications, this ability is of crucial practical importance. The computational cost of the dynamic simulations grows at least quadratically with the size of the grid. Using ML methods for predicting dynamic stability becomes useful for real-world application, if they can be trained on datasets of small synthetic networks, which are easy to simulate, while still performing well on large, complex real-world grids.

\subsection*{Ensembles of synthetic and real world power grid topologies modeled with Kuramoto oscillators}
\label{sec:method_details_on_datasets}

\begin{figure}
    \centering
    \includegraphics[width=\linewidth]{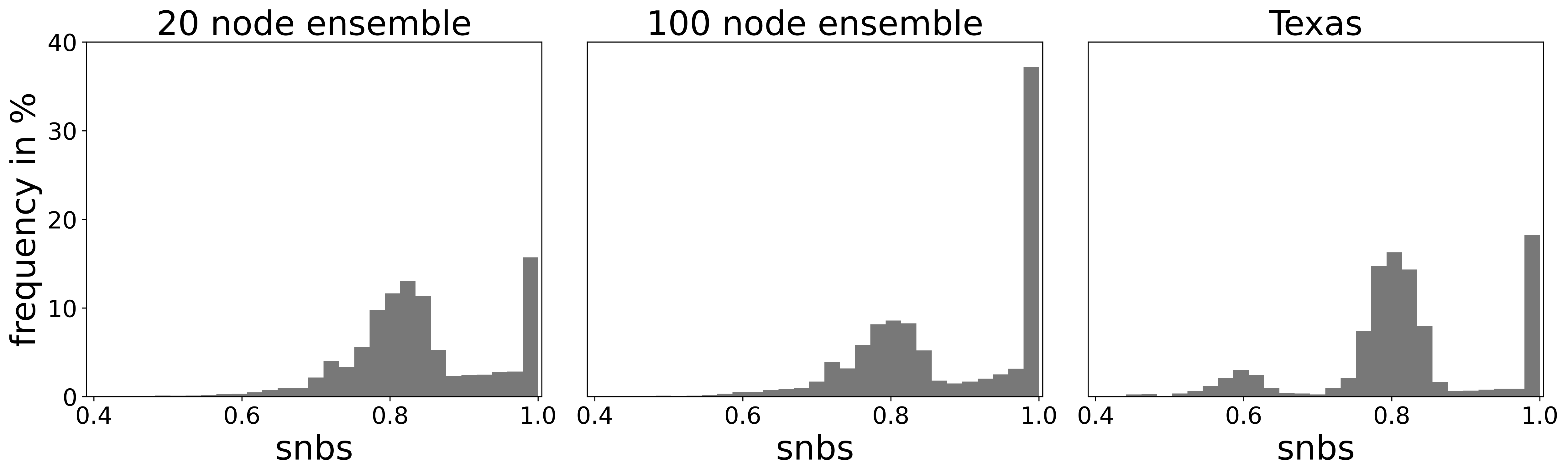}
    \includegraphics[width=\linewidth]{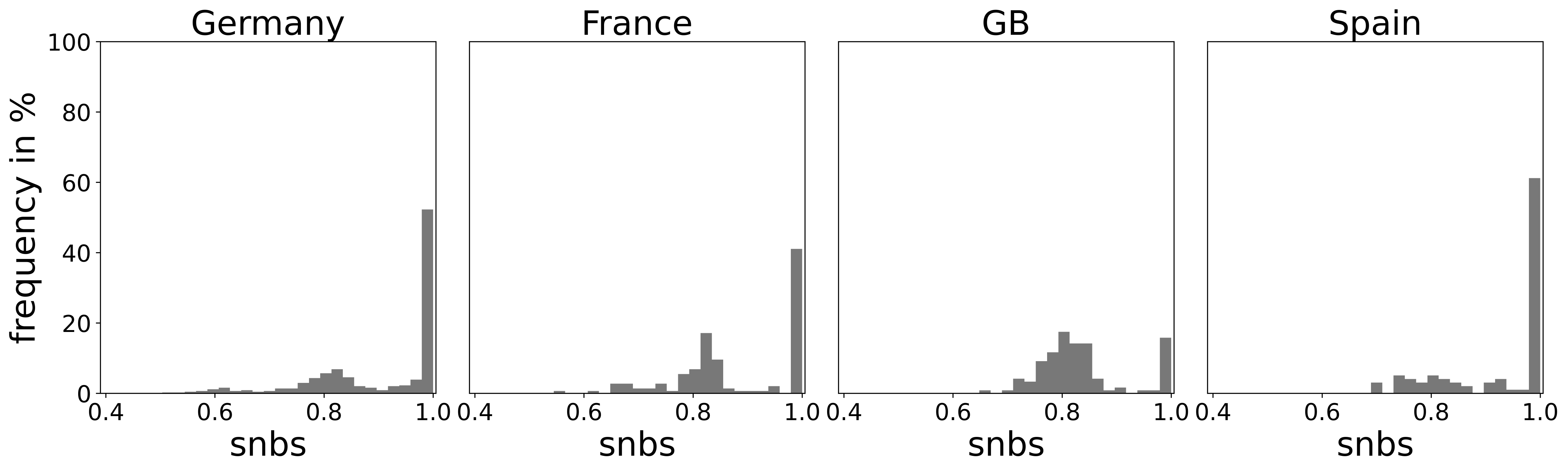}
    \caption{Histogram of SNBS for the synthetic ensembles and real grids. The 100 node ensemble has a higher proportion of completely stable nodes in comparison to the 20 node ensemble. For the Texas grid, a third mode of nodes with diminished stability appears, which poses challenges for ML models. There are only a few nodes with SNBS $<0.4$, hence we refrain from showing them. However, the minimum SNBS among all nodes is 0.21 for the 20 node ensemble, 0.05 for 100 node ensemble and 0.45 for the Texan grid. All the real topologies show a maximum SNBS of 1, but there are significant differences in the minima, which are for France: 0.374, for Germany: 0.297 for GB: 0.66, and for Spain: 0.69.}
    \label{fig:hist_snbs}
\end{figure}

To compare statistical differences regarding the dynamic stability of the real topologies in comparison to the synthetic datasets, we visualize the histograms of SNBS in \Cref{fig:hist_snbs}. The differences in distribution are caused by the differences regarding their generation. The grids of the 20 and 100 node ensembles are generated using a spatial random growth model \cite{schultzRandomGrowthModel2014} tuned to infrastructure and power grid properties \cite{schultzRandomGrowthModel2014}. The Texan grid is generated by a process that aims to ensure good electrical properties of the generated grid \cite{birchfieldGridStructuralCharacteristics2017,birchfieldACTIVSg20002000busSynthetic2021}.
Lastly, the real topologies have their own unique characteristics. The German grid, which is the largest among these four, has a small third peak for low SNBS values which is similar to the synthetic Texan grid. This might indicate that the three modes are not a unique property of the Texan grid, but a property that only emerges for grids of certain sizes. The GB grid is less stable than the other topologies. Compared to the other real and synthetic grids, there are considerably fewer nodes that are stable against all perturbations. This suggests systematic differences in the structure of the GB grid and the other grids. Therefore, we expect the GB grid to be particularly challenging for extrapolation tasks.

%
The datasets also contain alternative prediction targets related to dynamic stability, notably the maximum frequency deviation throughout the transient. In contrast to SNBS, NetSciML predicts this measure at high accuracy and generalizes well to other ensembles.

\subsection*{Complexity and training details of the used ML models}
GNN models require considerable training time, especially when conducting hyperparameter studies. Training of NetSciML is significantly faster. The NetSciML approaches based on gradient-boosted trees and linear regression have negligible training times of less than 2 CPU minutes. 

By far the most significant computational effort is required for the hyperparameter studies of the GNNs. For this analysis, it is common to analyze hundreds or thousands of possible configurations. In contrast, hyperparameter studies considerably less resource-intensive for gradient boosting trees (GBT). This underscores the significant advantage of employing the NetSciML approach, which incurs substantially lower computational costs.
Once the models are trained, ML approaches have the huge advantage of negligible evaluation times. This is especially helpful when evaluating unknown grids of the size of the Texan grid. For such a large grid, the required Monte-Carlo simulations take roughly 15,000 CPU hours \cite{nauckDynamicStabilityAssessment2023} when aiming for an accuracy comparable to current GNNs, whereas the evaluation takes only a few seconds for ML methods. 

\paragraph{Training and model details}


We use the trained The TAGNet model from  \cite{nauckDynamicStabilityAssessment2023}. The TAG model has has 13 layers and consists of 415,320 parameters. The number of channels is optimized per layer and varies between 30 and 141. No dropout is used.

\Cref{tab:PG-Kura-Asymptotic_gnn_std,tab:PG-Kura-Transient_gnn_std} report the performances across different model initializations.

\begin{table}[htbp]
\centering
\caption{PG-Kura Asymptotic: GNN $R^2$ scores reported as mean $\pm$ standard deviation across multiple random initializations.}
\label{tab:PG-Kura-Asymptotic_gnn_std}
\begin{tabularx}{\linewidth}{l>{\centering\arraybackslash}X>{\centering\arraybackslash}X>{\centering\arraybackslash}X>{\centering\arraybackslash}X>{\centering\arraybackslash}X>{\centering\arraybackslash}X>{\centering\arraybackslash}X}
\toprule
Model & 20 & 100 & DE & FR & GB & ES & TX \\
\midrule
GNN & {\cellcolor[HTML]{2171B5} \color[HTML]{F1F1F1} 0.716 $\pm$ 0.006} & {\cellcolor[HTML]{08519C} \color[HTML]{F1F1F1} 0.883 $\pm$ 0.001} & {\cellcolor[HTML]{08519C} \color[HTML]{F1F1F1} 0.911 $\pm$ 0.002} & {\cellcolor[HTML]{08519C} \color[HTML]{F1F1F1} 0.908 $\pm$ 0.004} & {\cellcolor[HTML]{4292C6} \color[HTML]{F1F1F1} 0.608 $\pm$ 0.017} & {\cellcolor[HTML]{08519C} \color[HTML]{F1F1F1} 0.838 $\pm$ 0.014} & {\cellcolor[HTML]{08519C} \color[HTML]{F1F1F1} 0.833 $\pm$ 0.015} \\
\bottomrule
\end{tabularx}
\end{table}

\begin{table}[htbp]
\centering
\caption{PG-Kura Transient: GNN $R^2$ scores reported as mean $\pm$ standard deviation across multiple random initializations.}
\label{tab:PG-Kura-Transient_gnn_std}
\begin{tabularx}{\linewidth}{l>{\centering\arraybackslash}X>{\centering\arraybackslash}X>{\centering\arraybackslash}X>{\centering\arraybackslash}X>{\centering\arraybackslash}X>{\centering\arraybackslash}X>{\centering\arraybackslash}X}
\toprule
Model & 20 & 100 & DE & FR & GB & ES & TX \\
\midrule
GNN & {\cellcolor[HTML]{08306B} \color[HTML]{F1F1F1} 0.956 $\pm$ 0.001} & {\cellcolor[HTML]{08306B} \color[HTML]{F1F1F1} 0.979 $\pm$ 0.000} & {\cellcolor[HTML]{08306B} \color[HTML]{F1F1F1} 0.996 $\pm$ 0.000} & {\cellcolor[HTML]{08306B} \color[HTML]{F1F1F1} 0.993 $\pm$ 0.001} & {\cellcolor[HTML]{08306B} \color[HTML]{F1F1F1} 0.993 $\pm$ 0.002} & {\cellcolor[HTML]{08306B} \color[HTML]{F1F1F1} 0.995 $\pm$ 0.002} & {\cellcolor[HTML]{08306B} \color[HTML]{F1F1F1} 0.956 $\pm$ 0.005} \\
\bottomrule
\end{tabularx}
\end{table}


\FloatBarrier

\section{Supplementary material on PG-Real experiments}
\label{sec:appendix_experiments_PGReal}

The PG-Real dataset consists of synthetic topologies with validated dynamical properties, comprising an ensemble of networks with 70 to 80 nodes and individual networks of 128, 256, 512, and 1024 nodes. The dynamics are modeled using universal inverter-based resource models that accurately capture real power grid behavior. Stability metrics include transient stability measured by fault ride-through probability and asymptotic stability measured by the probability of returning to narrow frequency and voltage bounds.



The correlation between network measures and dynamical properties shows considerable variation (\Cref{tab:NF all correlations,tab:NF abs correlations,tab:NF var correlations}. Overall correlation values are low, with the notable exception of dynamic node types and transient stability. The heterogeneous dynamical properties appear more directly influential than topological characteristics. For transient stability prediction, GNNs demonstrate excellent performance (\Cref{tab:NF hgbr gnn}), while NetSciML lagg significantly. Predicting asymptotic stability proves much more challenging, achieving only modest $R^2$ values of $0.33$ and $0.16$ on the 70 to 80 node ensemble (\Cref{fig:results_combined_netsciml_vs_gnn}). However, GNNs maintained relatively consistent performance when scaling to larger grids, whereas NetSciML fail on substantially on larger networks. This pattern aligns with observations from the PG-Kura experiments. \Cref{tab:NF hgbr data efficiency} reports the performance depending on the number of training samples.

\subsection*{Details on analyzing realistic power grids with inverters}
\label{sec:supplement_inverter_powergrids}
Building on the work of \cite{nauckPredictingFaultRideThroughProbability2026}, we perform simulations on larger power grids to evaluate the generalization capabilities of our models across varying grid sizes. We follow the same algorithms as outlined in the original paper, with one key modification during the grid generation process. Specifically, for larger grids, it was challenging to identify a valid operating point. To address this, we initially treated the lines as static to obtain a feasible operating point before reintroducing the dynamic aspects of the actual power grids.

\paragraph{Training and model details}
The GNN used is a two-layer DBGNN, with each layer comprising 30 internal steps. The number of hidden channels is set to 120 for both nodes and edges. In total, the model contains 147,961 parameters. For the transient task, node and edge dropout are set to 0.014877 and 0.002454, respectively. For the asymptotic task, dropout is set to 0.1 at both node and edge level. 

One-cycle learning-rate schedulers are used with the following settings: initial division factor $413{,}759$ (transient) and $4{,}365{,}873$ (asymptotic); final division factor $1{,}241{,}798{,}872{,}312.98$ (transient) and $3{,}680{,}955{,}167{,}359.520508$ (asymptotic); and maximum learning rate $863.4449718452441$ (transient) and $851.067199$ (asymptotic).

\Cref{tab:PG-Real-Asymptotic_gnn_std,tab:PG-Real-Transient_gnn_std} show mean performance with standard deviation over model initializations.

\begin{table}[htbp]
\centering
\caption{PG-Real Transient: GNN $R^2$ scores reported as mean $\pm$ standard deviation across multiple random initializations.}
\label{tab:PG-Real-Transient_gnn_std}
\begin{tabularx}{\linewidth}{l>{\centering\arraybackslash}X>{\centering\arraybackslash}X>{\centering\arraybackslash}X>{\centering\arraybackslash}X>{\centering\arraybackslash}X}
\toprule
Model & 70-80 & 128 & 256 & 512 & 1024 \\
\midrule
GNN & {\cellcolor[HTML]{08306B} \color[HTML]{F1F1F1} 0.967 $\pm$ 0.001} & {\cellcolor[HTML]{08306B} \color[HTML]{F1F1F1} 0.953 $\pm$ 0.006} & {\cellcolor[HTML]{08306B} \color[HTML]{F1F1F1} 0.951 $\pm$ 0.010} & {\cellcolor[HTML]{08519C} \color[HTML]{F1F1F1} 0.907 $\pm$ 0.010} & {\cellcolor[HTML]{08519C} \color[HTML]{F1F1F1} 0.905 $\pm$ 0.006} \\
\bottomrule
\end{tabularx}
\end{table}

\begin{table}[htbp]
\centering
\caption{PG-Real Asymptotic: GNN $R^2$ scores reported as mean $\pm$ standard deviation across multiple random initializations.}
\label{tab:PG-Real-Asymptotic_gnn_std}
\begin{tabularx}{\linewidth}{l>{\centering\arraybackslash}X>{\centering\arraybackslash}X>{\centering\arraybackslash}X>{\centering\arraybackslash}X>{\centering\arraybackslash}X}
\toprule
Model & 70-80 & 128 & 256 & 512 & 1024 \\
\midrule
GNN & {\cellcolor[HTML]{9ECAE1} \color[HTML]{000000} 0.334 $\pm$ 0.035} & {\cellcolor[HTML]{C6DBEF} \color[HTML]{000000} 0.269 $\pm$ 0.144} & {\cellcolor[HTML]{4292C6} \color[HTML]{F1F1F1} 0.630 $\pm$ 0.056} & {\cellcolor[HTML]{DEEBF7} \color[HTML]{000000} 0.080 $\pm$ 0.125} & {\cellcolor[HTML]{DEEBF7} \color[HTML]{000000} 0.186 $\pm$ 0.047} \\
\bottomrule
\end{tabularx}
\end{table}

\section{Supplementary material on ML methods}
\label{sec:app_MLMethods}

\subsection*{Graph Neural Network architectures}
\label{appendix_GNN}
Different types of GNN have been introduced, some of which are detailed below. For the composition of the GNN models, we closely followed \cite{nauckPredictingBasinStability2022}. Many GNN layers build on the Graph Convolution Network (GCN) introduced by \cite{kipfSemiSupervisedClassificationGraph2017}:
\begin{equation}
	H = \sigma(\overline{A} X \Theta),
\end{equation}
where $H$ is the output of a layer, $\sigma$ denotes the activation function, using the input features $X$, the matrix $\Theta$ containing the learnable weights and a slightly modified and re-normalized adjacency matrix $\overline{A}$. To increase the considered region and to consider neighbors at further distance, multiple GCN layers can be applied consecutively. 

To increase the potential region per layer, \cite{duTopologyAdaptiveGraph2017}  use multiple exponents $i$ of $\tilde{A}$ within one layer according to the following scheme:
\begin{equation}
	H = \sum_{z=0}^Z D^{-\frac{1}{2}} A^z D^{-\frac{1}{2}} X \Theta_z.
\end{equation}

In case of DBGNN \cite{nauckDiracBianconiGraph2024}, both node and edge features are updated simultaneously using the following idea for the updates:
\begin{align}
\mathbf{h}_v^{(l+1)} = \sigma \left( \sum_{e \in \mathcal{E}(v)} \mathbf{W}_1 \mathbf{h}_e^{(l)} + \mathbf{W}_2 \mathbf{h}_v^{(l)} \right), \\
\mathbf{h}_e^{(l+1)} = \sigma \left( \sum_{v \in \mathcal{V}(e)} \mathbf{W}_3 \mathbf{h}_v^{(l)} + \mathbf{W}_4 \mathbf{h}_e^{(l)} \right),
\end{align}
where $\mathbf{h}_v^{(l)}$ and $\mathbf{h}_e^{(l)}$ are the node and edge features at layer $l$, $\mathcal{E}(v)$ is the set of edges incident to node $v$, $\mathcal{V}(e)$ is the set of nodes incident to edge $e$, and $\mathbf{W}_i$ are learnable weight matrices.

\subsection*{Best subset selection using orthogonal matching pursuit}
\label{sec:app_best_subset_selection}
To analyze the performance of small models with just a few independent variables, we perform best subset selections for the linear models.
Given input features $X_1,\ldots, X_p \in \mathbb{R}^n$ and target variables $y \in \mathbb{R}^n$, the best subset selection problem is to find parameters $\beta \in \mathbb{R}^n$, s.t.,
\begin{equation}
	\min_\beta \|y - (X\beta)\|^2_2 - \| \beta \|_0,
  \label{secBestSubsetSelection}
\end{equation}
with $\|.\|_0$ denoting the number of non-zero coefficients of the argument (sometimes called $\ell_0$-`norm'). This is just the least squares formulation of a linear model, with an added  regularization term.
As an algorithm for the best subset selection problem we use orthogonal matching pursuit (OMP) \cite{mallatMatchingPursuitsTimefrequency1993}, which approximates the subset of the network measures , that best fits the variance in the data. If the reduced model captures the main information and achieves good performance, it allows for a more intuitive understanding of the model predictions. We approximate the subsets 3 predictors and denote the corresponding models as OMP3.

\FloatBarrier
\printbibliography
\clearpage

\end{document}